\documentclass[useAMS,usenatbib]{mn2e}

\usepackage{graphicx}

\def\deg{$^{\circ}$}
\def\solm{M$_{\odot}$}
\def\solmyr{M$_{\odot}$yr$^{-1}$}
\def\kms{km s$^{-1}$}
\def\kmskpc{km s$^{-1}$ kpc$^{-1}$}
\def\etal{{et al.\ }}

\title[Nuclear spirals. II. Hydrodynamical models]
{Nuclear spirals in galaxies: gas response to asymmetric potential. 
II. Hydrodynamical models}

\author[Witold Maciejewski]{Witold Maciejewski\\
Obserwatorium Astronomiczne Uniwersytetu Jagiello\'{n}skiego,
ul. Orla 171, 30-244 Krak\'{o}w, Poland}

\begin{document}

\maketitle

\begin{abstract}
Nuclear spirals naturally form as a gas response to non-axisymmetry
in the galactic potential, even if the degree of this asymmetry is
very small. Linear wave theory well describes weak nuclear spirals, 
but spirals induced by stronger asymmetries in the potential are clearly 
beyond the linear regime. Hydrodynamical models indicate spiral shocks in 
this latter case that, depending on how the spiral intersects the $x_2$ 
orbits, either get damped, leading to the formation of the nuclear ring,
or get strengthened, and propagate towards the galaxy centre. 
Central massive black hole of sufficient mass can 
allow the spiral shocks to extend all the way to its immediate vicinity, 
and to generate gas inflow up to 0.03 \solmyr, which coincides with the 
accretion rates needed to power luminous local Active Galactic Nuclei.
\end{abstract}

\begin{keywords} 
hydrodynamics --- shock waves --- galaxies: kinematics and dynamics ---
galaxies: ISM --- galaxies: spiral --- galaxies: structure --- galaxies:
nuclei --- ISM: kinematics and dynamics
\end{keywords} 

\section{Introduction}
It has been recently proposed that nuclear spirals in galaxies may be
related to the fueling of Seyfert activity (Regan \& Mulchaey 1999). This 
was a straightforward conclusion when a search for the fueling mechanism using
highest-resolution optical observations of a sample of Seyfert nuclei 
with the Hubble Space Telescope returned nuclear spirals in 6 out of 12
galaxies. Nuclear spirals turned out to be much more frequent in Seyfert
galaxies than gas inflow related to nuclear bars, the commonly proposed
feeding mechanism. Observations of a larger sample that followed (Martini
\& Pogge 1999) found nuclear spirals in 20 out of 24 Seyfert 2 galaxies,
and in a later sample of 46 Seyfert 1 and 2 galaxies (Pogge \& Martini 2002) 
almost all were classified as having nuclear spirals. The authors of these 
latter surveys also showed that nuclear spirals are not self-gravitating, 
and that they are likely to be shocks in nuclear gas discs. 
The most recent study (Martini et al. 2003a,b) involves a sample of 64 Seyfert 
galaxies, as well as a control sample, which together are big enough so that
trends can be noticed. In particular, the authors of this study point out that 
all grand-design nuclear spirals occur in barred galaxies, but not all barred
galaxies develop nuclear spirals --- some of them have nuclear rings.
Tightly-wound nuclear spirals tend to avoid barred galaxies instead. 

\begin{table*}
\caption{The list of models}
\begin{tabular}{rcrrrrl} \\
\hline
\#& model& MBH mass& sound speed    &  type of   & radial extent & inner grid\\
  & name & in \solm& in gas        &  asymmetry & of the grid   & boundary \\
\hline
1 & 0W20o&    0    & 20 km s$^{-1}$ & weak oval  & 0.02 - 16 kpc & outflow\\
2 & 0W05o&    0    &  5 km s$^{-1}$ & weak oval  & 0.02 - 16 kpc & outflow\\
3 & 8W20o&  $10^8$ & 20 km s$^{-1}$ & weak oval  & 0.02 - 16 kpc & outflow\\
4 & 8W20r&  $10^8$ & 20 km s$^{-1}$ & weak oval  & 0.02 - 16 kpc & reflection\\
5 & 8W20c&  $10^8$ & 20 km s$^{-1}$ & weak oval  & 0.005 - 4 kpc & reflection\\
6 & 8W05o&  $10^8$ &  5 km s$^{-1}$ & weak oval  & 0.02 - 16 kpc & outflow\\
7 & 0S20r&    0    & 20 km s$^{-1}$ & strong bar & 0.02 - 16 kpc & reflection\\
8 & 0S05r&    0    &  5 km s$^{-1}$ & strong bar & 0.02 - 16 kpc & reflection\\
9 & 8S20r&  $10^8$ & 20 km s$^{-1}$ & strong bar & 0.02 - 16 kpc & reflection\\
10& 0D20o&    0    & 20 km s$^{-1}$ & double bar & 0.02 - 16 kpc & outflow\\
\hline
\end{tabular}
\end{table*}

Why do some barred galaxies develop nuclear spirals, while other develop 
nuclear rings? Why do tightly wound nuclear spirals prefer galaxies that do not
have a bar? Can any type of nuclear spirals generate inflow sufficient to feed 
local Seyfert nuclei? Seyfert galaxies require mass accretion rates of 
$\sim 0.01$ \solm/yr (e.g. Peterson 1997).
Here I attempt to answer these questions under an assumption that
nuclear spirals are density waves generated in gas by a rotating potential,
as described in the accompanying Paper I (Maciejewski 2004). Implications
of the linear theory (originally proposed by Goldreich \& Tremaine 1979)
derived in Paper I will serve here as a guideline, but by themselves they 
cannot provide answers to the questions above, since the amplitude of
strong bars very much exceeds the linear theory.

In the linear approximation the arm/interarm density ratio is a scalable
value, as long as the perturbation is small. To estimate how big this ratio
is for a given asymmetry in the potential, one can search for nonlinear
solutions (e.g. Yuan \& Cheng 1989, 1991) or directly involve hydrodynamical 
modeling. The second approach has the advantage that a whole range of 
non-axisymmetries, from small ones to ones of the order of the axisymmetric 
component, can be studied with the same tool. I take this approach, and I 
construct in this paper hydrodynamical models of gas flow in the nuclear 
regions of weakly and strongly non-axisymmetric potentials. 

To study structures in gaseous nuclear discs of galaxies with the help of 
hydrodynamical models exceptionally high resolution is required, which has
been achieved only recently. Athanassoula's models of gas flow in barred
galaxies (1992b) show curling of the inner parts of the straight principal 
shock, 
but the resolution of these models prevents us from following this feature 
further inwards. Nuclear spirals generated by the bar inside the straight 
principal shocks, and winding by more than a $\pi$-angle,
were first noticed in hydrodynamical simulations by 
Maciejewski (1998, 2000), and Englmaier \& Shlosman (2000). The latter work
interpreted these features in terms of spiral density waves, weak enough
so that the linear theory is applicable. On the other hand, Maciejewski,
Teuben, Sparke \& Stone (2002, hereafter MTSS02) point out that nuclear 
spirals in their models take the form of a shock, which is beyond the 
scope of the linear treatment.

In this paper I construct hydrodynamical models of nuclear spirals for
realistic gravitational potentials represented by rotation curves 
characterized in Paper I. In particular, I am interested in how the gas flow
in the nucleus is modified by the presence of a central massive black hole
(MBH) or a density cusp. As pointed out in Paper I, the central MBH 
significantly changes the nuclear gravitational potential, and therefore it
should be able to regulate gas flow around itself. Here, I investigate how its
presence modifies gas inflow onto the centre. In Section 2, I list the models
to be analyzed in this paper, and I describe the code with which they were 
built. In order to link the models with the linear theory, in Section 3 I 
analyze models of gas flow in a weak oval, where they should not depart
significantly from the linear prediction. In Section 4, I apply the same
analysis to gas flow in strong bars. Preliminary results about nuclear spirals 
in double bars are listed in Section 5.

\section{The code and setup of the models}
The models were calculated using the CMHOG hydrodynamical code (Piner et al.
1995, MTSS02), which solves the single-fluid equations in
their Eulerian form on a fixed polar grid. The gas is isothermal with the
sound speed $c=20$ km/s (hereafter termed hot gas), the value suitable 
for centres of galaxies (Englmaier \& Gerhard 1997), but runs with 
$c=5$ km/s (hereafter termed cold gas) have also been done for comparison.
The gas is not self-gravitating. All models are built on the grid covering
half of the plane, and point symmetry is assumed. The grid has 174 cells
spaced logarithmically in the radial direction, and 80 cells covering a
180\deg\ angle.

In order to discuss the physical processes operating in relation to nuclear 
spirals, I chose 10 hydrodynamical models showing typical features for a
given gravitational potential and gas characteristics. The models, with their 
parameters, are listed in Table 1. 
Models starting with '0' are built for the potential characterized by 
rotation curve $A$ from Paper I (linear inner rise), and models starting with 
'8' are for the potential characterized by rotation curve $B$ (a $10^8$ \solm\ 
MBH added in the centre). The potential in all models 
includes an $n=2$ Ferrers bar. The bar is either identical to the primary bar
in models of MTSS02 (termed 'strong bar' in Table 1, model names with 'S'), or 
its quadrupole moment is 10 times smaller than in MTSS02 models, and the axial 
ratio of the bar is decreased to 1.5 from the original 2.5 for strong bar 
(hereafter I call it 'weak oval', model names with 'W'). 
The potential with a double bar is identical to that in MTSS02.

In the linear approach, the potentials of both the strong bar and the weak
oval have the outer Inner Lindblad Resonance (ILR) at 2.3 kpc, and, in the 
absence of a central MBH (models starting with '0'), they also have an 
inner ILR (iILR) at 0.13 kpc. Models with a central MBH (models starting 
with '8') have no iILR. 

For each model, the initial gas density is constant throughout the grid (10 
\solm\ pc$^{-2}$), and the initial kinematics is gas motion on circular orbits 
with rotation velocity derived from the axisymmetrized potential, where the 
bar mass was incorporated into the bulge component. Then, through the
first 0.1 Gyr of each run, the bar or oval is extracted continuously
from the bulge, and its strength remains unchanged afterwards till the
end of the simulation. The method to introduce the secondary bar is
discussed in Section 5 devoted to models of nested bars.

Polar grid has singularity at $r=0$, and models built on it cannot include 
the galactic centre, but they stop at a certain minimal radius: the inner grid
boundary. The calculated gas flow in the innermost parts of the galaxy may
depend on the boundary conditions adopted there. Usually outflow conditions
are imposed with no inflow onto the grid allowed (e.g. Piner et al. 1995,
Englmaier \& Shlosman 2000, MTSS02). However, this boundary condition 
effectively means creation of a sink for gas, which may generate unphysical 
inflow, and unclear rules for wave reflection or absorption. Therefore in 
this paper I introduced also a reflection condition at the inner grid 
boundary. The benefit of this boundary condition is that the sink term
is removed from the problem, because no gas leaves the grid. Consequently,
more conservative estimates for gas inflow can be given (Section 3.4).
With this condition, waves in gas are fully reflected at the inner boundary.
In Section 3.3 I show that the reflected wave is unlikely to play important
role in gas dynamics.

\section{Gas flow in a weak oval}
Models 1--6 are built for the potential with a weak oval, whose departure 
from axisymmetry is much smaller than that for a strong bar studied by MTSS02. 
The $Q_T$ parameter, defined as the maximum ratio of tangential to radial force
(Combes \& Sanders 1981), is 0.21 for
models with a strong bar, but only 0.01 for models with a weak oval. In a 
real galaxy, such asymmetry will most likely remain undetected, leading to
unbarred classification even in the recent detailed infrared studies (e.g.
Laurikainen, Salo \& Buta 2004).

\subsection{Global morphology and kinematics of a nuclear spiral in hot gas}
A snapshot of gas density, representative for a nuclear spiral generated
by a weak oval in hot gaseous disc is shown in Fig.1. It was taken from 
Model 8W20, 0.4 Gyr into the run, once the flow has stabilized. Model 0W20
shows the same morphology, except for its innermost parts, because
gravitational potentials in these two models are almost identical at radii
above a few hundred parsecs. The nuclear spiral is clearly visible in 
the top panel of Fig.1, although the straight principal shocks disappeared 
completely once the bar was replaced by a weak oval. The density contrast 
between the arm and the inter-arm region is about 2 (Fig.1, middle panel), 
therefore the spiral should be clearly visible in the color maps. On the 
other hand, this contrast is small enough, so that the perturbation of the 
velocity field is small. Therefore there are no shocks forming 
in the gas, and gas flow in discs with this type of the nuclear spiral is 
almost circular.

\begin{figure}
\centering

\vspace{-27mm}

\includegraphics[width=0.95\linewidth]{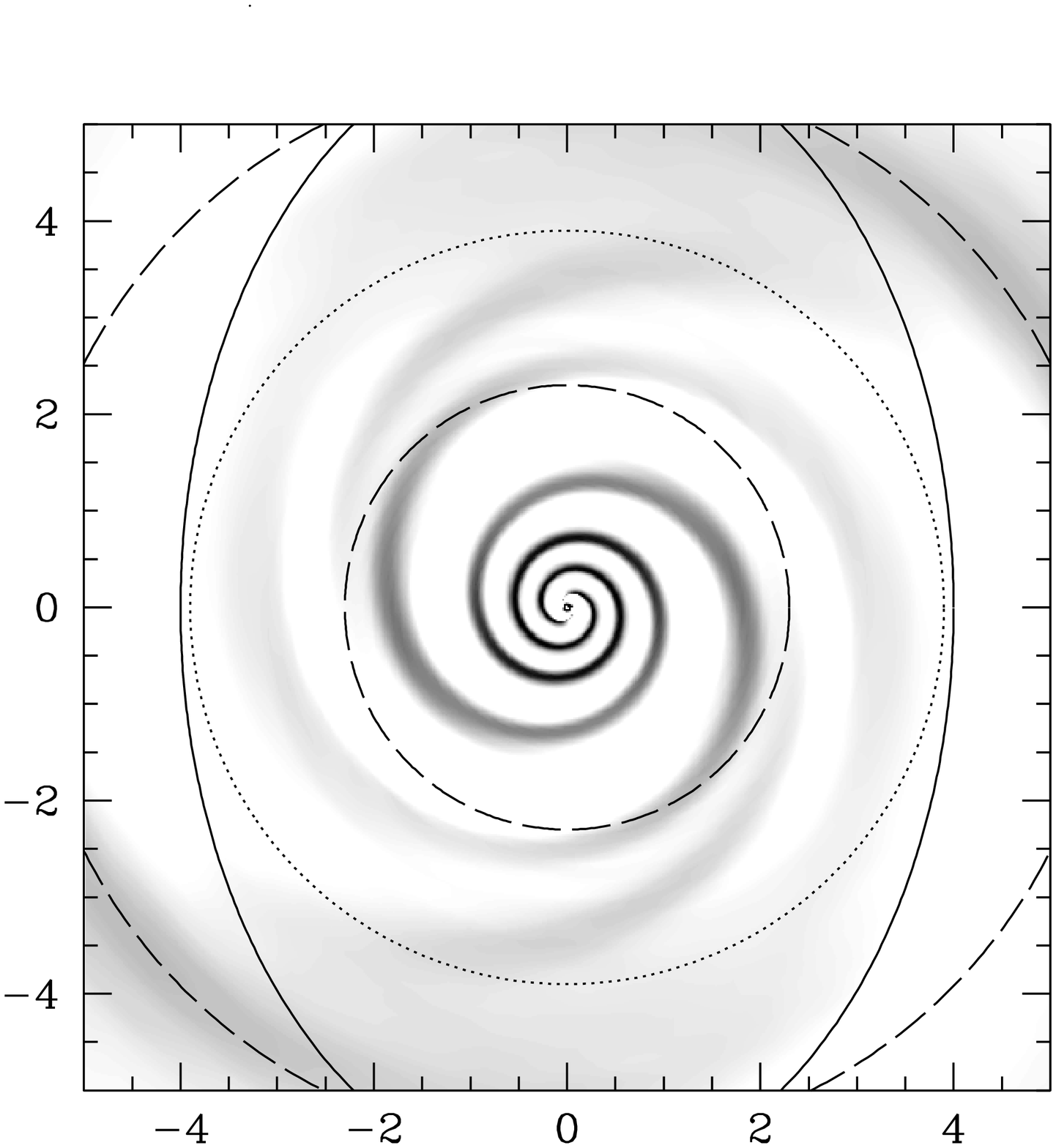}

\vspace{-5mm}

\includegraphics[width=0.72\linewidth]{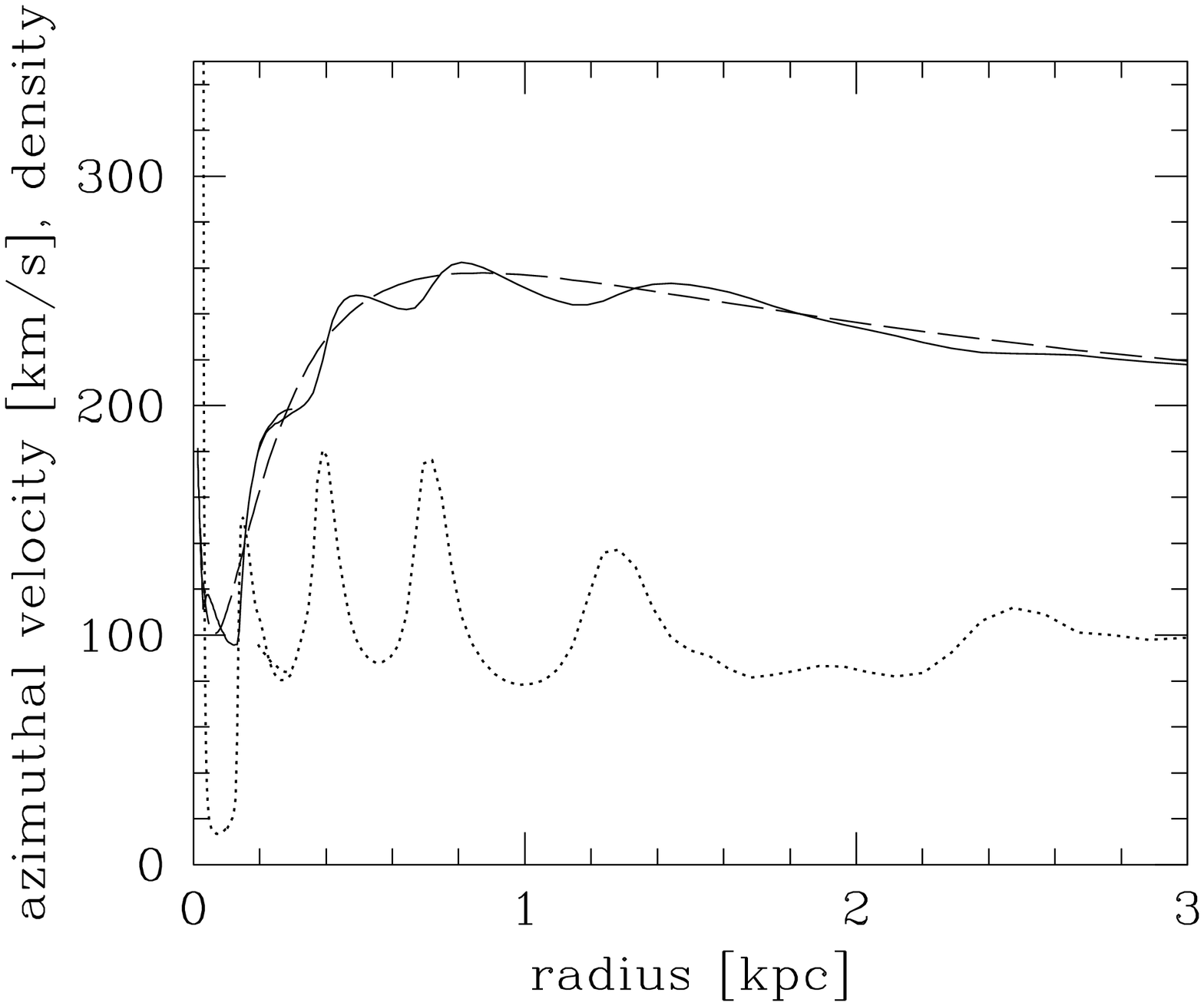}

\vspace{-12mm}

\includegraphics[width=0.72\linewidth]{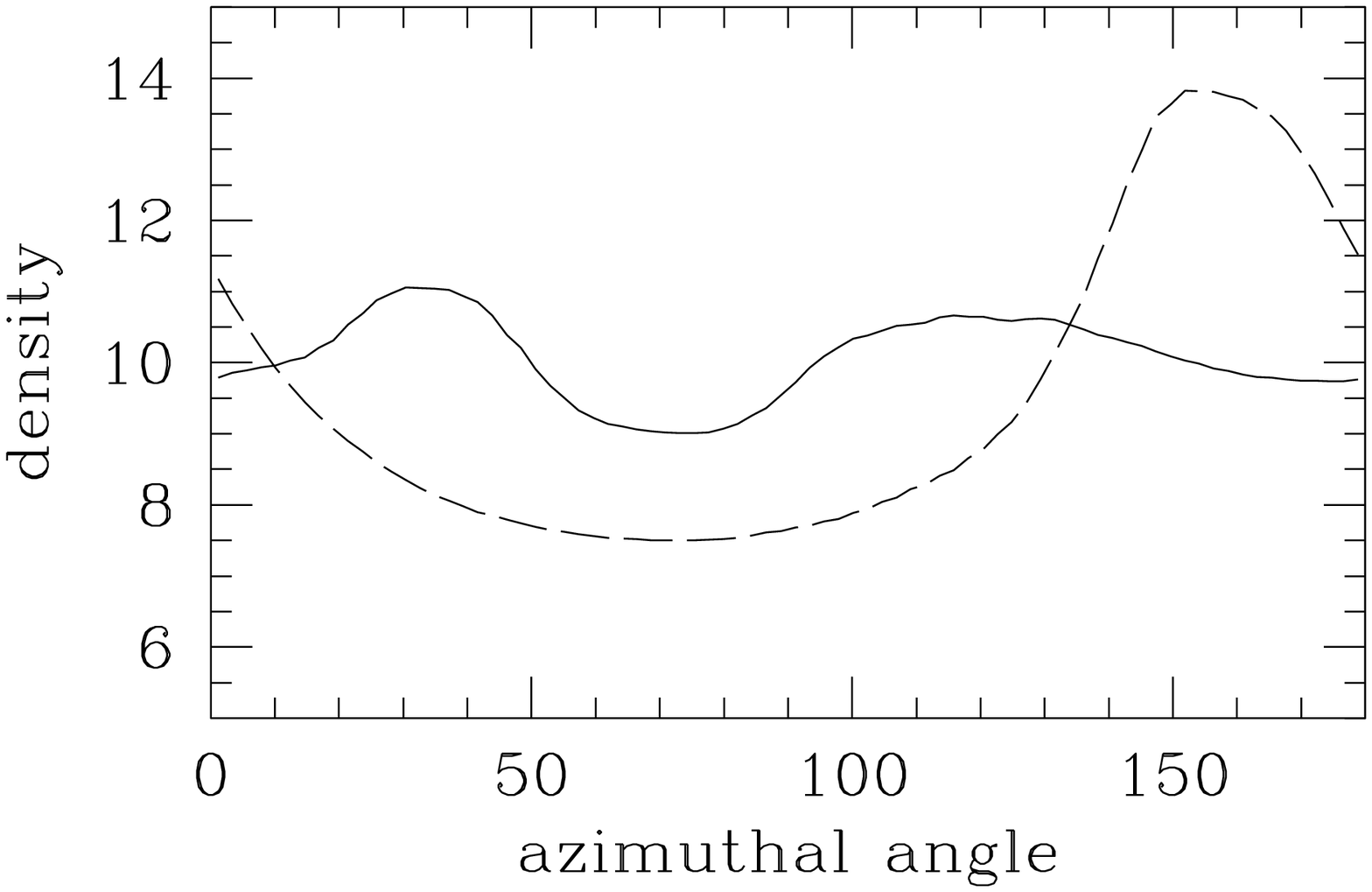}

\vspace{-20mm}

\caption{{\bf Top:} Snapshot of gas density in model 8W20r, in which
a weak oval asymmetry is present in the potential characterized by 
rotation curve $B$ in Paper I. Gas density is shown
at time 0.4 Gyr, after the morphology of the flow has stabilized. Darker 
color indicates larger densities. The solid ellipse outlines the oval, and the
dashed circles mark the oILR at 2.3 kpc and the corotation at 5.6 kpc. The 
dotted circle marks the position of the 4:1 resonance at 3.9 kpc. Units on 
axes are in kpc.
{\bf Middle:} Radial density profile (dotted line), and azimuthal velocity 
(solid line), plotted against rotation curve $B$ from Paper I (dashed line)
as a function of radius along the vertical line in the top panel, in models 
8W20, at time 0.4 Gyr. To show the structure of the innermost regions, at
radii smaller than 0.3 kpc data from model 8W20c are being used instead of
8W20r. The velocity units are in \kms, the density units are arbitrary.
{\bf Bottom:} Density variation in model 8W20r along two circles: at 1.5 kpc 
(dashed line) and at 3 kpc (solid line). Because of the assumed bisymmetry 
of the models, variations over only 180\deg\ are shown. The density
units are arbitrary, the azimuthal angle is in degrees.}
\label{f3}
\end{figure}

Unlike nuclear spirals in strong bars, which unwind outwards rapidly in 
order to match the principal shock in the bar, nuclear spirals in weak 
ovals follow the linear mode longer, and wind up to 3 times around the 
centre (6$\pi$ angle). After the flow gets stabilized, the bisymmetric
nuclear spiral seen in Fig.1 remains unchanged in the frame rotating 
with the bar: it does not rotate, neither it winds up or unwinds around
the centre. Between the inner grid boundary, and the outer ILR (oILR) it winds 
around the centre by about a 5$\pi$ angle.

\begin{figure*}
\centering
\vspace{-45mm}
\includegraphics[width=1.05\linewidth]{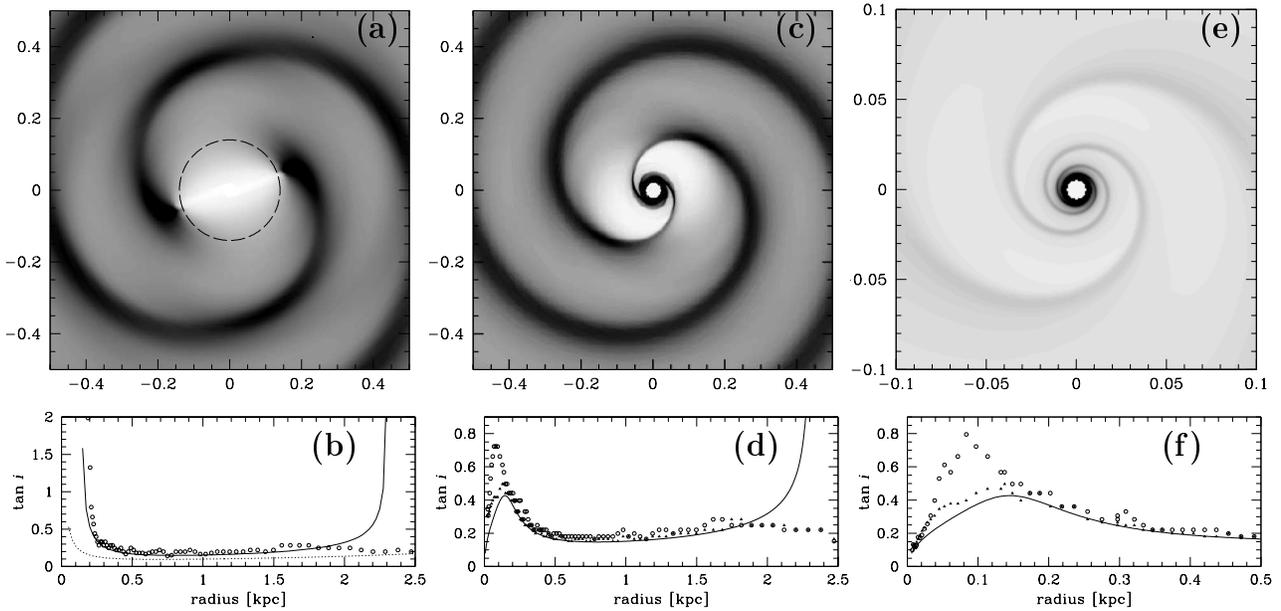}
\vspace{-115mm}
\caption{{\bf a)} Snapshot of gas density in the innermost regions of
model 0W20o at time 0.4 Gyr, after the flow has stabilized. The dashed
circle marks the iILR, units on axes are in kpc. The major axis of the oval 
is vertical here, and in all consecutive figures.
{\bf b)} Open circles mark tangens of the pitch angle $i$ of the nuclear 
spiral in model 0W20o at 0.4 Gyr, solid line is the linear prediction 
for an $m=2$ spiral, and dotted line indicates the prediction for an $m=4$ 
spiral.
{\bf c)} Same as {\it a)}, but for model 8W20r.
{\bf d)} Same as {\it b)}, but for model 8W20r. Filled triangles mark pitch 
angle measured for the hydrodynamical model at 250 Myr, before the flow 
stabilizes.
{\bf e)} Same as {\it a)}, but for the innermost 100 pc of model 8W20c,
at 250 Myr, when the winding nuclear spiral is present.
{\bf f)} Same as {\it d)} but for model 8W20c.}
\label{f4}
\end{figure*}

\subsection{The $m=4$ spiral outside the oILR}
In the linear approximation, the nuclear spiral should not extend outwards 
beyond the outer ILR (Paper I, Section 3.1). In fact, from the top panel of
Fig.1 one can see that a clear double-arm spiral does not extend out beyond 
the oILR at 2.3 kpc (dashed circle). However, there {\it is} 
spiral structure detectable in gas morphology out to about 3.5 kpc. 
The density contrast is much weaker there, and a closer inspection
indicates that a four-arm spiral is present outside the two-arm one.
Linear theory (see Paper I, Section 3) predicts that such a spiral can be 
generated by an $m=4$ mode in the potential, and that it should 
extend from the galactic centre out to the radius where $\Omega - \kappa/4 
= \Omega_B$. This is the $(4:-1)$ resonance in notation of Paper I, hereafter 
called for simplicity 4:1. For our potential it is located at 
3.9 kpc (dotted circle in the top panel of Fig.1), which is consistent with 
the observed extent of the four-arm spiral.
Ferrers' bar can be decomposed into even-$m$ components, among them $m=4$, 
and this component is responsible for a four-arm spiral outside the proper 
nuclear spiral. Note that two arms of this four-arm spiral are just 
continuations of the two-arm nuclear spiral from smaller galactic radii,
albeit with much lower density contrast. In addition, two other arms start 
at the oILR, at position angles $\sim 90$\deg and $\sim 270$\deg, and
extend outwards. The transition from a two-arm spiral inside the oILR to 
a four-arm one outside it is illustrated in the bottom panel of Fig.1, 
which shows the density 
profiles as a function of angle along the rim of two circles: one of radius 
1.5 kpc, which is located inside the oILR, and the other one of radius
3 kpc, placed between the oILR and the 4:1 resonance. Along the first
circle, there is only one clear density maximum per $\pi$ angle, which
indicates a two-arm spiral. The density ratio is about 1.8. Along the
second circle, two weak but still clear density maxima are seen in a
$\pi$ angle. This is characteristic for a four-arm spiral. The density 
ratio between maxima and minima along this circle ranges between 1.1
and 1.2, depending on which maximum/minimum values are taken.
Thus one may expect weak four-arm spiral structures outside grand-design
two-arm nuclear spirals.

The smooth transition from the nuclear spiral to the four-arm spiral in
the hydrodynamical models can be also seen in the radial changes of the 
pitch angle (Fig.2b, open circles). It closely follows the linear prediction 
for a two-arm spiral inside the oILR (Fig.2b, solid line), but its value 
remains almost unchanged also at larger radii, where the nuclear spiral should 
unwind and disappear. There, spiral arms of the nuclear spiral continue 
outwards, assuming outside the oILR pitch angle predicted by the linear theory
for a four-arm spiral (dotted line in Fig.2b). Thus continuation 
of the nuclear spiral to larger radii may hide the presence of the oILR in 
the galaxy.

\subsection{Morphology of the innermost regions}
The linear theory says that if inside the oILR another ILR is present, 
the nuclear spiral should not propagate inwards of this ILR (the iILR). 
Here I follow the hydrodynamical realization of this rule, using models of 
gas flow in gravitational potentials with the iILR (models 0W20), and without
it (models 8W20). Note that the inclusion of a $10^8$ \solm\ MBH, which is the
sole difference between the gravitational potentials in these models, is
sufficient to remove the iILR in a galaxy with a constant-density core.

In models 0W20 (rotation curve $A$), the 
nuclear spiral unwinds rapidly when it approaches the iILR from the outside, 
with its pitch angle well following the linear prediction (Fig.2b), and it 
disappears just outside the iILR. It remains strong all the way until 
reaching the iILR, which may be the reason why the leading spiral predicted
by linear theory to form at the iILR (Paper I) is absent.
On the other hand, the 8W20 models (rotation curve $B$) have only one ILR, 
and a clear nuclear spiral extends there all the way to the inner boundary 
(Fig.2c).

\begin{figure}
\centering
\vspace{-12mm}
\includegraphics[width=0.48\linewidth]{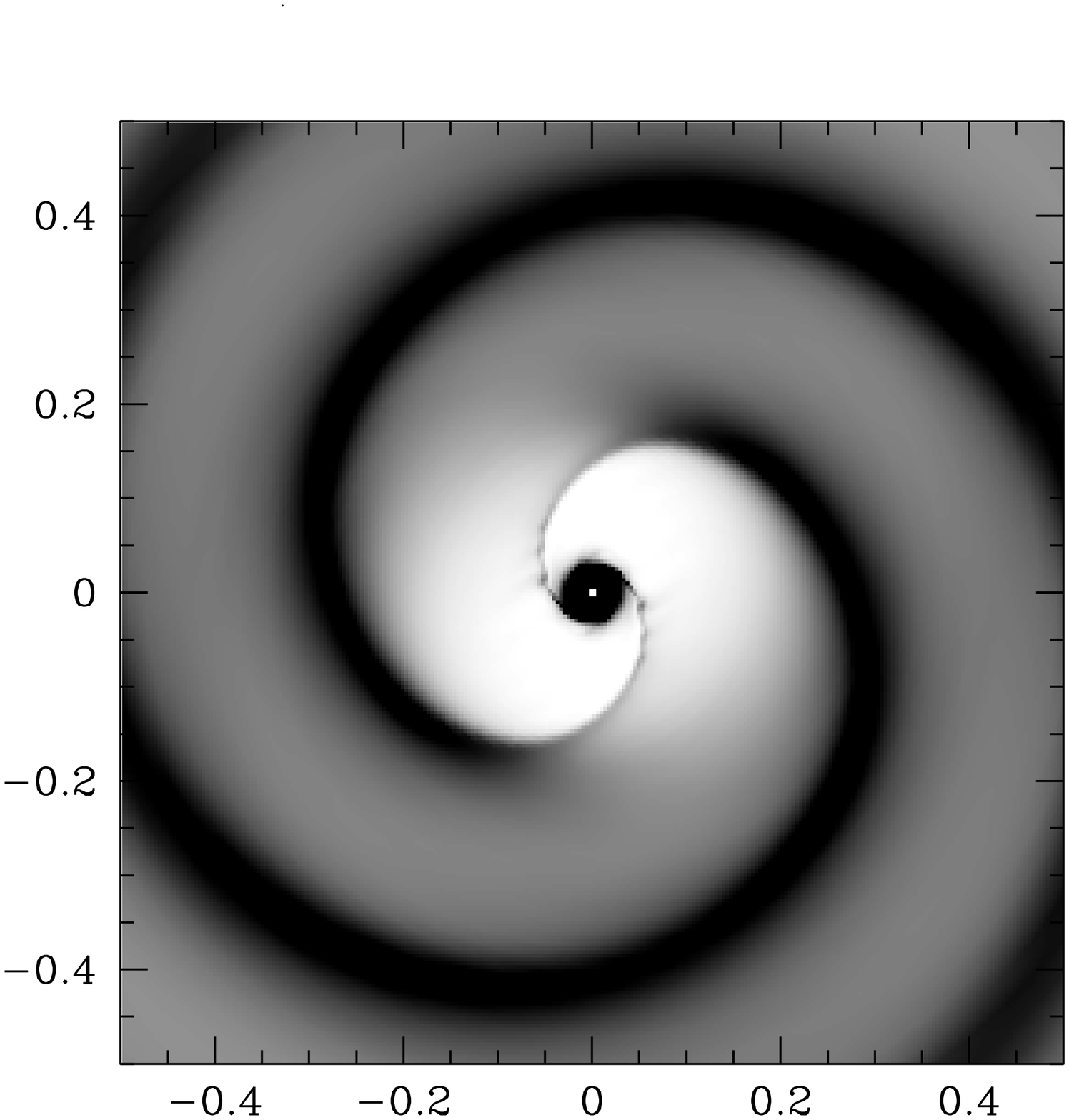}
\includegraphics[width=0.48\linewidth]{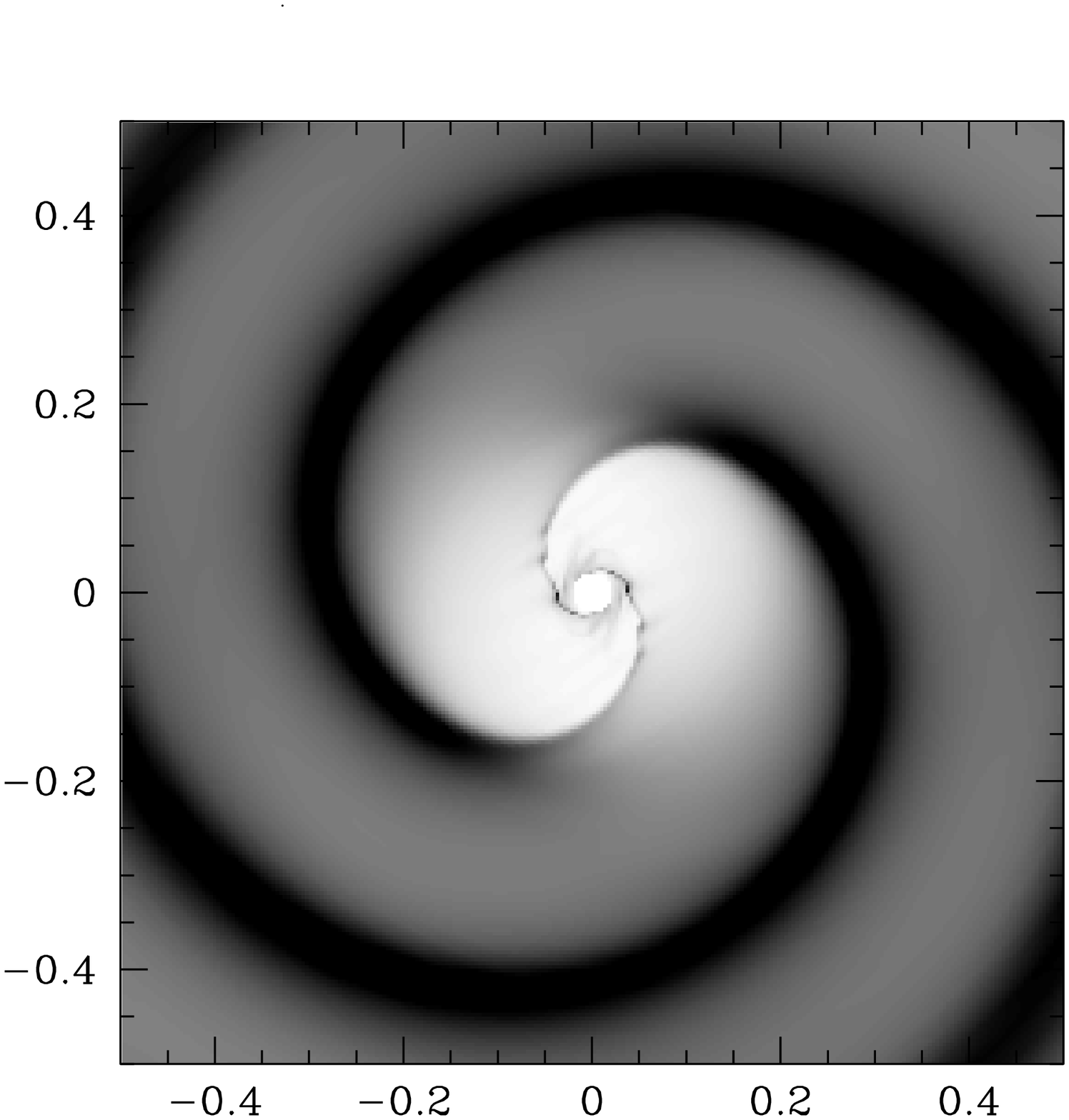}
\vspace{-5mm}
\caption{Density distribution in the innermost parts of models 
8W20c ({\it left}), and 8W20o ({\it right}) at the end of the run at 0.9 Gyr.}
\label{f5}
\end{figure}

In the model 8W20r, the wave generating this nuclear spiral reflects from
the inner boundary and interferes with the incoming wave, which may perturb 
the solution. However, the original wave moving inwards is focused towards
the centre, while the reflected wave diverges away from the centre, and it is
quickly overcome by the incoming wave. Thus the perturbation caused by 
reflection does not propagate beyond the innermost 10-20 cells, which on
the standard grid corresponds to the range of radii 30 -- 45 pc. Moreover, this
boundary condition has parallels to the actual physical situation, when
the wave propagating inwards encounters the accretion disc of the MBH with
density likely higher, which causes reflection, and any inflow in the spiral 
accumulates in the accretion disc.

Nevertheless the steady-state solution 
for model 8W20r does not reflect the winding of the spiral in the innermost 
regions predicted by the linear theory (compare the central-bottom panel of
Fig.2 in Paper I to Fig.2c here). This is clearly seen as a discrepancy 
between the value of the pitch angle predicted by the linear theory,
and measured in the model (Fig.2d, open circles) at radii below 200 pc.
After the flow stabilizes, the maximal pitch angle (36\deg) is 
clearly larger than the linear prediction (23\deg), and the maximum occurs at 
a smaller radius. However, when the model is examined before the flow 
settles down (here at 0.25 Gyr), the measured pitch angle is much closer
to the linear prediction (Fig.2d, triangles).

To investigate whether this effect is numerical (vicinity of the inner grid
boundary), I built two more versions of model 8W20: one with the outflow
inner boundary condition (8W20o), and one still with the reflective inner
boundary, but extending four times further towards the galaxy centre, down
to the radius of 5 pc (8W20c). In this last version, a nuclear spiral winding 
up towards the centre develops during the early stages of the simulation.
Its shape (Fig.2e) is similar to that in the central-bottom panel of Fig.2 
in Paper I, and its pitch angle closely follows the linear theory (Fig.2f, 
triangles). However, when the flow stabilizes, the innermost part of the 
nuclear spiral unwinds, with its pitch angle growing and reaching the same 
values as in model 8W20r (Fig.2f, open circles). This result remains unchanged
when outflow through the inner boundary is allowed (model 8W20o). In fact, 
once the flow stabilizes, the morphology of the nuclear spiral in all 3 
versions of model 8W20 gets identical, and it remains so till the end of 
each run at 0.9 Gyr (Fig.3). Regardless of whether gas accumulates in the
innermost cells of the grid in models with reflective inner boundary condition
(8W20c, 8W20r) or whether it is removed from the grid when outflow is allowed
(8W20o), the nuclear spiral reaches the same steady state.
Thus I conclude that the unwinding of the innermost part
of the nuclear spiral is not an effect of proximity of the inner grid 
boundary, but rather has hydrodynamical origin. However, it is unlikely
to be an effect of wave reflection in the galaxy centre, since it also
appears in the model version with the outflow boundary condition.

\subsection{Gas inflow triggered by a nuclear spiral in a weak oval}
The rate of inflow can be deduced from how the mass contained within various
radii changes with time. For model 8W20r, Fig.4 shows gas mass within a number 
of radii as a function of time. Only mass enclosed in the innermost
circle of the radius of 40 pc changes significantly. Throughout the run
it increases from 0.37$\times 10^5$\solm\ to 2.14$\times 10^5$\solm,
but most of the inflow occurs between 0.25 Gyr and 0.4 Gyr, when over
1.5$\times 10^5$\solm\ is dumped into radii below 40 pc. Thus
average inflow during this period of 150 Myr is about $10^{-3}$\solmyr.
This inflow occurs exactly when the transition from a tightly wound spiral 
(Fig.2e) to the steady-state solution (Fig.2c) occurs. Thus the formation
of a nuclear spiral results in a single event of gas inflow into 
innermost parsecs of the galaxy, which dumps there about $10^5$\solm\
of gas. After this single dump, the mass inflow is negligible, consistent
with zero. Such a one-time dump happens only in the models with the nuclear 
MBH (8W20). In models without it (0W20) the mass accumulated in the innermost
regions does not change significantly with time.

\begin{figure}
\centering
\includegraphics[width=0.9\linewidth]{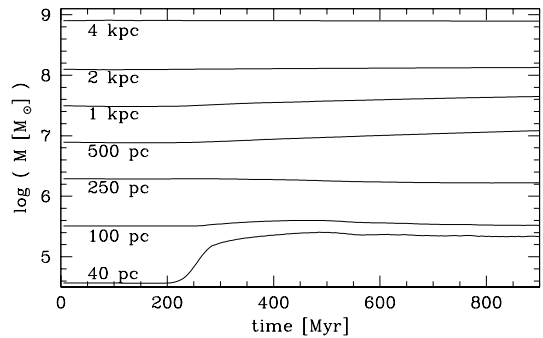}
\vspace{-28mm}
\caption{Mass accumulated within various radii (indicated in the plot)
as a function of time for model 8W20r.} 
\label{f6}
\end{figure}

Nuclear spirals may be generated not only by weak ovals, but also
by transient phenomena like a passing globular cluster or a giant molecular
cloud. Such nuclear spirals would then also be transient and reoccurent.
Model 8W20r indicates that every time the spiral reappears, it dumps some 
$10^5$\solm\ of gas onto the innermost parsecs of the galaxy, which may 
provide a way to sustain a weak nuclear activity. However, for such reoccurent
dumping to take place, material has to be replenished into the inner 100pc,
because formation of the spiral leads to mass increase within this radius of a 
few percent only, while 60\% of gas within 100 pc radius ends up within
40 pc radius after formation of the spiral.

\begin{figure}
\centering
\vspace{-27mm}
\includegraphics[width=0.95\linewidth]{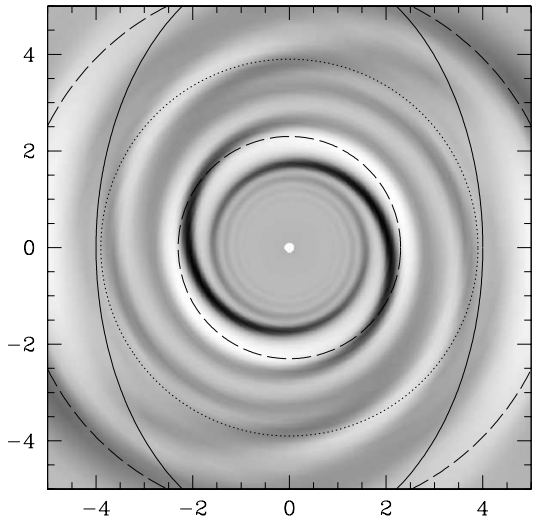}
\vspace{-5mm}
\caption{Snapshot of gas density in model 8W05o, in the same galactic
potential as in model 8W20r presented in Fig.1, and at at the same 
evolutionary time. The only difference is the assumed sound speed in the
isothermal gas, which is 5 \kms\ here (cold gas). Reference lines and the
scale are the same as in the top panel of Fig.1.}
\label{f7}
\end{figure}

Note that by imposing the reflection condition on the inner boundary of the 
polar grid used in these simulations I get the lower limit for the inflow. 
If free outflow through the inner boundary is imposed instead, 
(model 8W20o, as in Piner et al. 1995), a flux of about $2 \times 10^{-3}$ 
\solmyr is crossing this boundary continuously after the flow stabilizes. 
Although this can be interpreted as an inflow, it is likely caused by the 
assumption here that gas can leave, but not re-enter the grid, which creates 
a sink term in the problem.

The mass enclosed within larger radii does not change significantly
under the action of the nuclear spiral. There is some inflow at radii
between 0.5 and 1 kpc, since the mass enclosed within these radii increases
during the 0.9-Gyr run by 56\% and 42\% of its initial value, respectively.
The mass accumulated there increases gradually throughout the run, 
which translates into an average inflow up to 0.01 \solmyr. However, this
mass does not get much further inwards than 500 pc from the galaxy centre,
since the mass accumulated within the radius of 250 pc actually
slightly decreases throughout the run. Thus no mass transport from
kpc- to pc-scale is expected by nuclear spirals of this kind.

\subsection{Nuclear spirals in cold gas}
The linear theory predicts that the pitch angle $i$ of the spiral (see eq.24
in Paper I) is proportional to the speed of sound in the isothermal gas in 
which the density wave propagates. Thus the spiral in models 0W05 and 8W05, 
that involve cold gas with sound speed of 5 \kms, is much more tightly wound.
Only model 8W05 is shown (Fig.5), since the other one looks almost 
identical. From the linear theory, the pitch angle in cold-gas models is
expected to be four times smaller than in the hot-gas models. For a 
tightly-wound spiral, at any radius $R$, the radial distance $dR$ between the 
adjacent density maxima is $\pi R \tan i$. In cold-gas models considered here
it can be be as small as $0.13R$. This corresponds to the radial separation 
of only 4 cells on our grid, which is insufficient to resolve the waves and 
results in numerical damping. Such an effect is seen in Fig.5, where stronger
spiral is only present close to the oILR at 2.3 kpc, where its pitch angle
is expectedly larger. There, the arm-interarm density ratio approaches 3.
This amplitude gets damped quickly inwards, although the spiral
can be traced down to the radius below 1 kpc, where it winds almost
by a $6 \pi$ angle.

As in the hot-gas case, a four-arm spiral is seen outside the oILR. 
According to the linear theory (Paper I), a leading nuclear spiral should 
develop in the vicinity of the iILR in model 0W05o. In cold gas its 
propagation should not be affected by the 
interference with the trailing spiral originating at the
oILR, as it happens in hot-gas models. Interestingly, such a leading spiral 
does not form -- I discuss reasons for it in Section 6.3. Gas inflow in 
cold-gas models is
negligibly small: in both 0W05 and 8W05 it never exceeds $10^{-5}$\solmyr
even when outflow from the grid is allowed through the inner boundary.

\section{Nuclear gas flow generated by a strong bar} 
The perturbation in the stellar gravitational potential coming from 
a typical galactic bar is too strong to be described in linear
terms. Thus gas flows generated by such perturbation cannot be well 
described by the linear wave theory. One can get a better insight 
from the orbital theory of bars (Athanassoula 1992a, see also reviews
by Sellwood \& Wilkinson 1993 and by Maciejewski 2003). Hydrodynamical
models indicate that in the main body 
of the bar, two symmetric shocks (the principal shocks) form on the leading 
sides of the bar. If the bar is strong and if it extends to its own 
corotation, these shocks are straight. Otherwise the shocks curl and start 
resembling a trailing spiral (Athanassoula 1992b). If there is an ILR in 
the galaxy, the 
principal shocks do not point at the galactic centre, but they are offset 
from it. Gas and dust get compressed in the principal shocks, which 
are seen as dust lanes in the optical images of barred galaxies (see
e.g. NGC 1097, NGC 1300, NGC 4303, and NGC 6951 in the Hubble Atlas of 
Galaxies, Sandage 1961). Inside the inner ends of the dust lanes, there are 
nuclear rings (e.g. NGC 4314, Benedict et al. 2002) or nuclear 
spirals (e.g. NGC 1530, Regan, Vogel \& Teuben 1997, Pogge \& Martini 2002).

\begin{figure*}
\centering

\vspace{-27mm}

\includegraphics[width=0.48\linewidth]{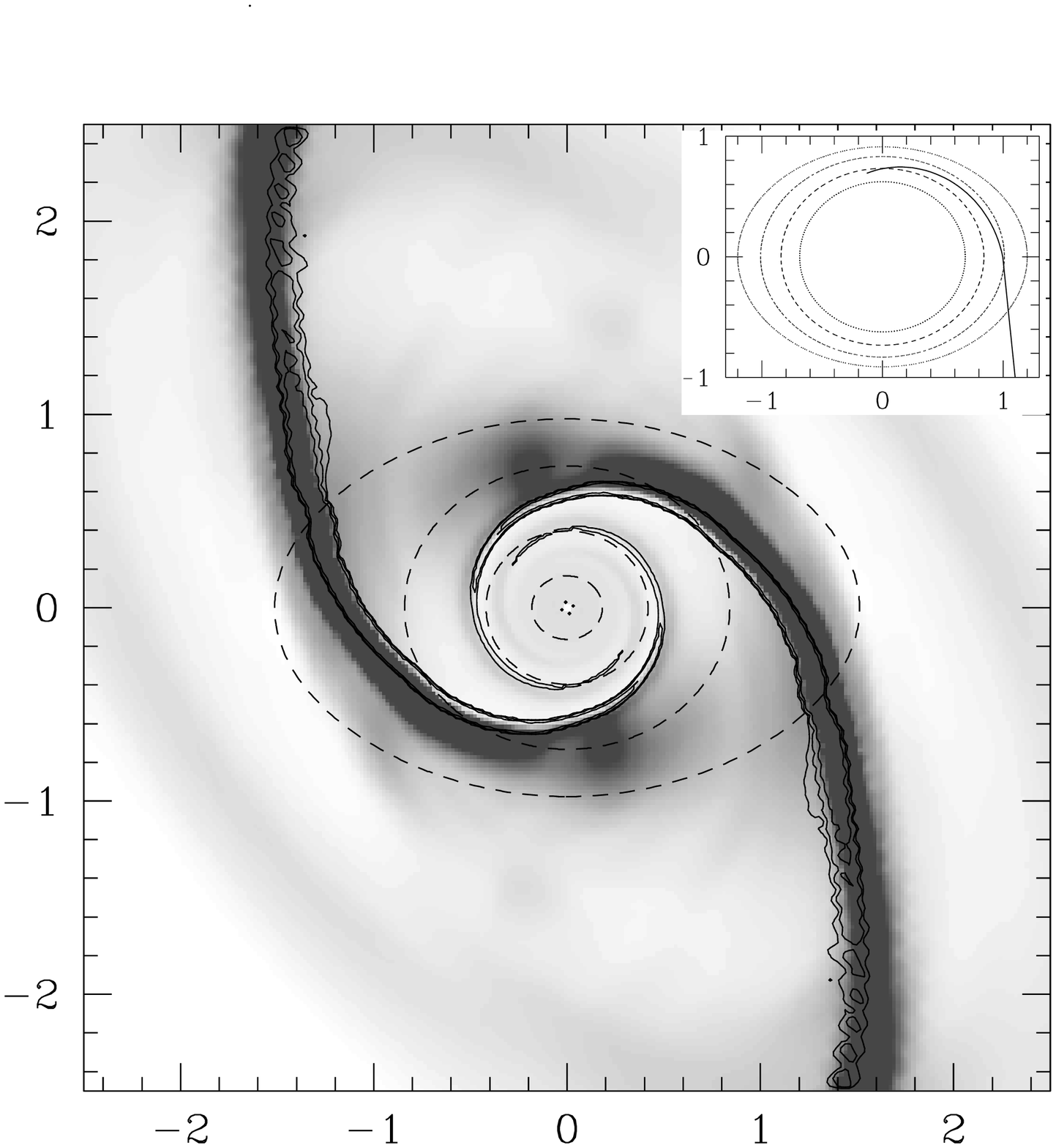}
\includegraphics[width=0.48\linewidth]{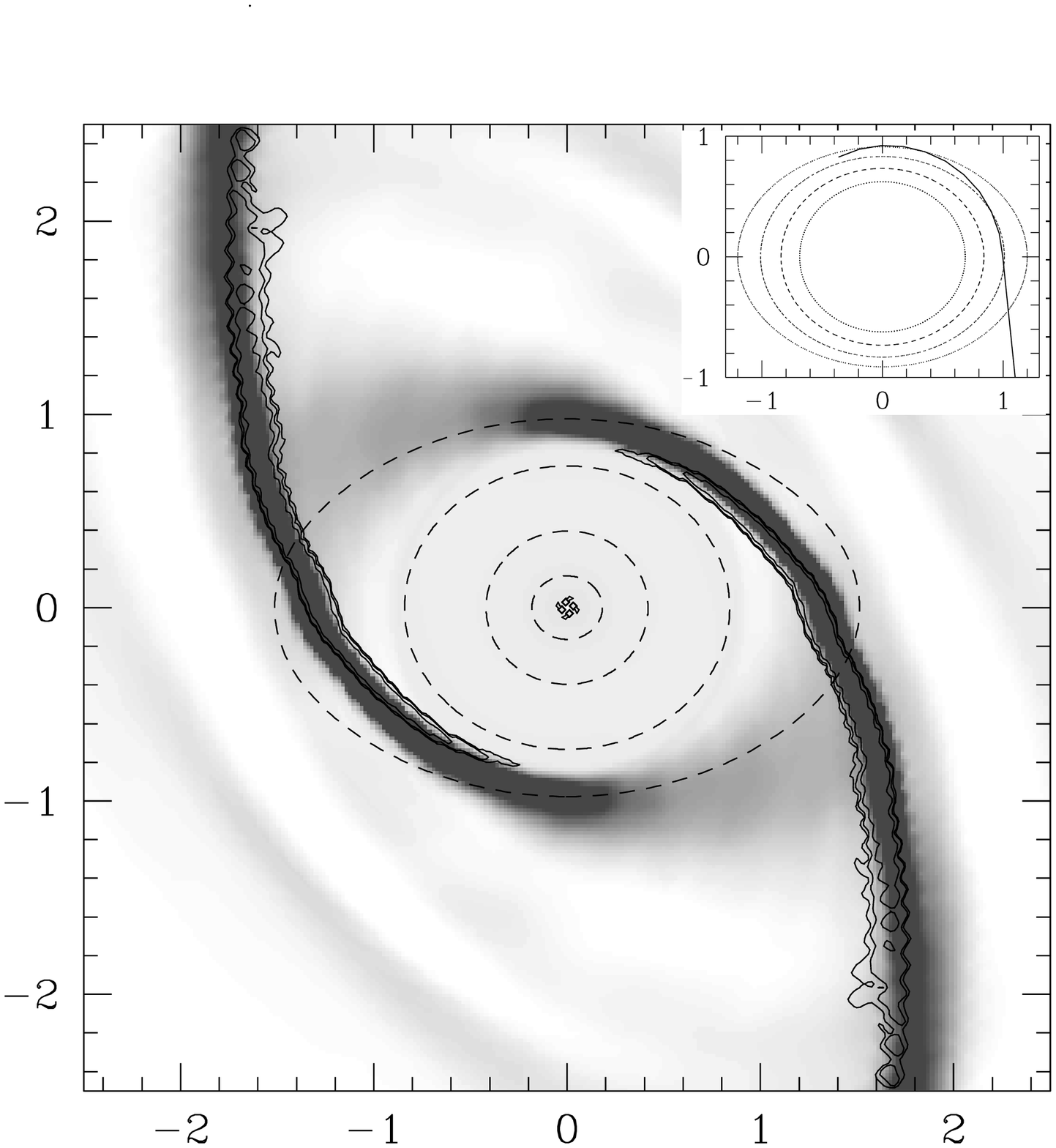}

\vspace{-5mm}

\caption{{\bf Left:} Snapshot of gas flow in a strong bar (vertical) in model 
0S20r (hot gas) at 110 Myr, when the nuclear spiral is being formed. Gas 
density is in 
greyscale, $div^2 {\bf v}$ for $div {\bf v} < 0$ (shock indicator) is in 
contours. Exemplary orbits from the $x_2$ family are drawn with dashed lines.
Note how the shock propagates out of the density enhancement, and spirals 
into the innermost regions. The insert shows the $x_2$ orbits (dotted lines)
together with the adjusted linear prediction
of the shape of the spiral (solid line) for the sound speed of the model,
appended by a straight line in the lower half of the plot that mimics the
principal straight shock.
{\bf Right:} Same for gas flow in model 0S05r (cold gas). Note how the shock 
propagating inwards in radius is damped by the density enhancement. Units on 
axes are in kpc.}
\label{f8}
\end{figure*}

Until recently, limited computing resources prevented us from studying
these nuclear structures inside the principal dust lanes, because the
resolution of the models was not high enough. In the models by Athanassoula
(1992b), the straight principal shocks curl inwards in their innermost parts, 
but they cannot be followed to much smaller radii because of the limited
resolution. Piner et al. (1995) employed polar grid in their hydrodynamical 
code (CMHOG, used also to build models presented in this paper). Resolution 
of this grid increases inwards, which allowed them to clearly 
resolve nuclear rings. They explained formation of these rings in terms of 
the orbital structure in the bars. This explanation is 
summarized in Section 4.1 below, where I compare the mechanisms leading to 
the formation of nuclear rings and spirals. 
The CMHOG code allowed also to resolve for the first time nuclear spirals 
inside the straight principal shocks in the bar (Maciejewski 1998, 2000). 
Similar nuclear spirals were seen by Englmaier \& Shlosman (2000) in
their models with a different code, but also on the polar grid. In this
section I analyze in detail gas flow in the central regions of a strongly 
barred galaxy.

\subsection{How to generate nuclear ring, and how nuclear spiral}
It has been commonly accepted that the nuclear ring in barred galaxies
forms when the shocked gas leaves the $x_1$ orbits and settles on the
lower-energy $x_2$ ones (Athanassoula 1992a,b, Piner et al. 1995). The 
$x_2$ orbits are almost round, and they do not intersect one another, 
making therefore a perfect location for gas to accumulate.

However, strong bar, like any asymmetry in the nuclear potential, 
should also generate nuclear spirals inside its ILR. In the orbital
theory, the ILR is defined as the outer limit to which the $x_2$ orbits
can extend. Thus the nuclear spiral should be cospatial
with the nuclear ring. Yet some barred galaxies show clear nuclear
rings, while other display nuclear spirals without rings. Still some
show nuclear spirals inside nuclear rings. What is the reason for this
variety?

Formation of the nuclear ring is explained by the orbital theory,
and the shapes of orbits that underlie this ring solely depend on the
properties of the gravitational potential. On the other hand, the nuclear
spiral has a wave-like nature, and its properties are determined by the
dispersion relation (eq.~16 in Paper I). This relation depends not only on 
the gravitational potential, but also on the gas characteristics. If gas is 
assumed to be isothermal, it depends on the sound speed in gas. 

Thus two mechanisms: orbital and wave-like, compete in shaping the dynamics 
of gas flow in central parts of galaxies. Using two models of gas flow in a 
strongly barred galaxy, I examine the outcome of this competition when the 
gas is hot (model 0S20r), and when it is cold (model 0S05r). Aside for two 
different sound speeds in gas, all other parameters in these models are
identical.

On the early stages of evolution, right after the bar has reached its full 
strength, the shocked gas tends to settle on the $x_2$ orbits marked in 
Fig.6 with dashed lines. At the same time, the inner parts of the principal 
shock in the bar curl inwards, and tend to follow the linear dispersion 
relation inside the ILR. Thus in general, the shocked gas tends to follow
a path different from the direction in which the shock propagates, because
the pitch angle of the the shock is different from the pitch angle of the 
orbit at a given location (Fig.6, inserts). Moreover, the linear formula 
for the pitch angle of the spiral wave (eq.~24 in Paper I) indicates that 
this angle is larger for larger sound speeds.

In the hot gas (model 0S20r, Fig.6, left panel), the sound speed is high, 
thus the pitch angle of the spiral is large. As shown in the insert, this
pitch angle is {\it always larger} than the pitch angle of the $x_2$ orbits 
which the shock crosses. The shock propagates inwards crossing each $x_2$ 
orbit only once, and then moving to smaller orbits. Thus the post-shock gas,
which tends to settle on these orbits, always moves away from the shock front. 
In other words, the spiral shock propagates {\it out of} the post-shock gas 
condensation, into regions where the density of gas is much lower. 
This is clearly seen in the left panel of Fig.6, after the shock crosses 
the second outermost $x_2$ orbit. Propagation of shock from a high-density to
low-density medium triggers a shoe-lace effect: the shock gets strengthened. 
In model 0S20r it gains strength high enough that it continues to propagate 
all the way to the galactic centre. On the other hand, the post-shock gas 
tends to settle on the $x_2$ orbits, as the gas condensation between the two 
outermost $x_2$ orbits indicates.

In the cold gas (model 0S05r, Fig.6, right panel), the sound speed is low, 
thus the pitch angle of the spiral is small. It can become smaller than the 
pitch angle of the $x_2$ orbits at their intersection with the shock (Fig.6, 
insert of the right panel). When it happens, the 
shock, still propagating inwards, crosses the $x_2$ orbits from inside out.
This means that the dispersion relation forces the shock to propagate back
{\it into} the post-shock gas, which tends to settle on these $x_2$ orbits. 
In this gas condensation the shock gets damped. This happens before
the shock reaches the major axis of the bar in the right panel of Fig.6.
Once the spiral shock weakens and disappears inwards, the gas settles on 
closed trajectories originating from the $x_2$ orbits. However, they are not 
exactly the $x_2$
orbits, since the shock constantly penetrates the gas lane from inside,
forcing the steady-state solution to be rounder than the $x_2$ orbits in 
the same area. This mechanism is confirmed by detailed hydrodynamical 
simulations of nuclear rings, where they appear almost circular, while
the underlying $x_2$ orbits are significantly flattened (Piner et al. 1995,
MTSS02).

In short, the nuclear spiral shock, being a continuation of the straight
principal shock in the bar, can propagate towards the galactic centre 
when it is able to escape the gas condensation emerging from this principal 
shock. It can do it when the pitch angle of the spiral is large enough.
If this pitch angle is too small, the spiral shock gets damped in this gas
condensation, and a nuclear ring forms.

\begin{figure}
\centering

\vspace{-27mm}

\includegraphics[width=0.95\linewidth]{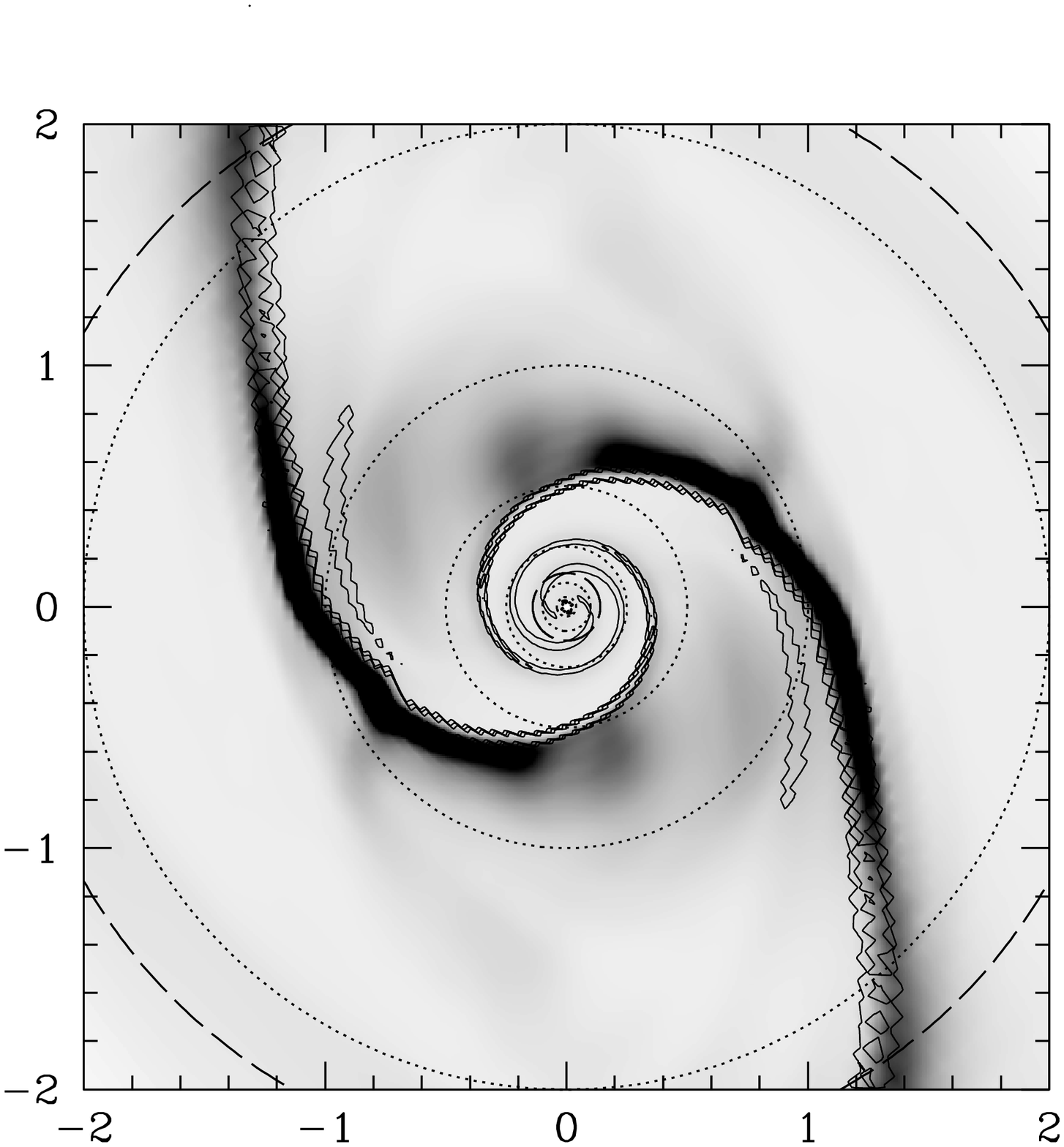}

\vspace{-4mm}

\includegraphics[width=0.75\linewidth]{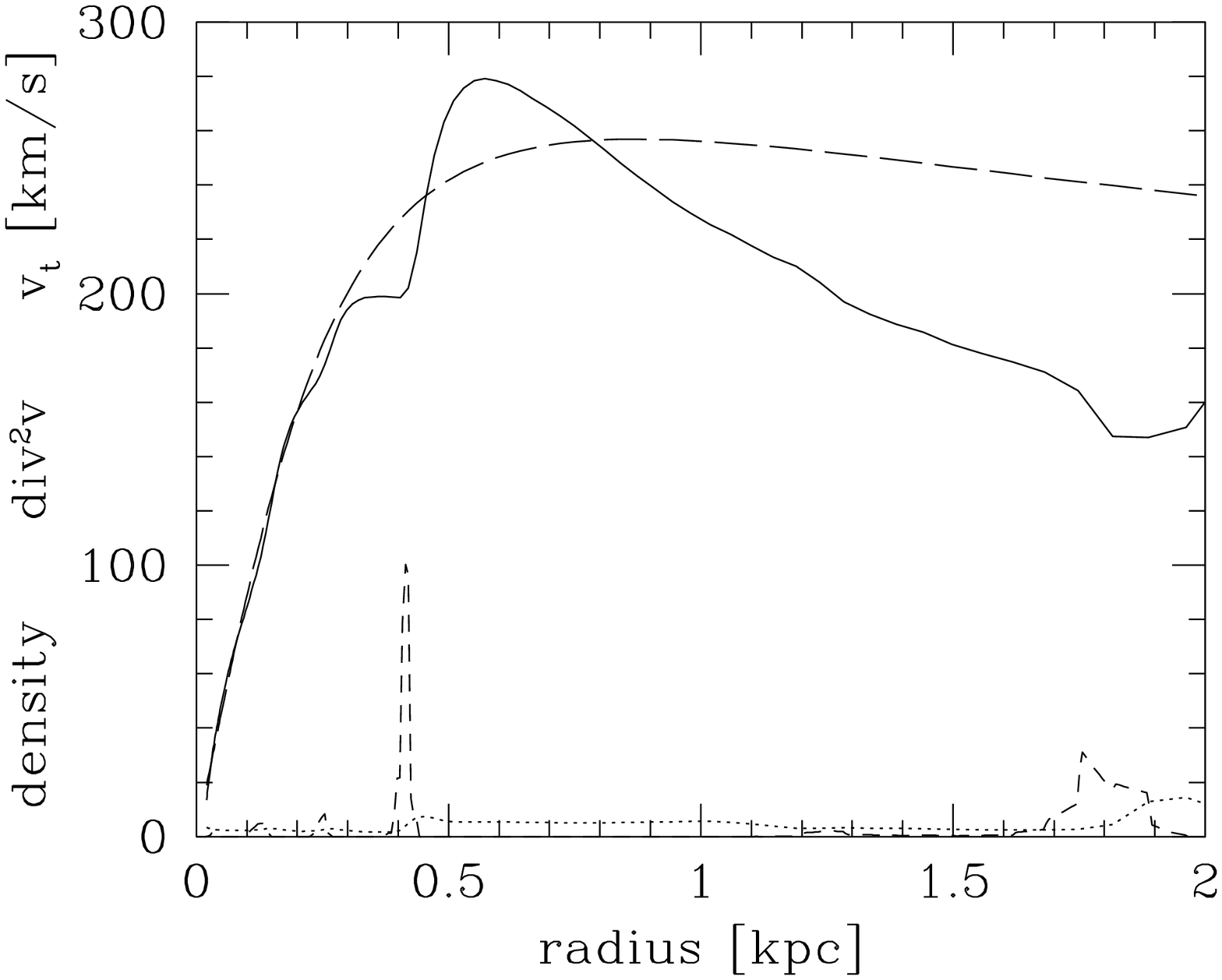}

\vspace{-14mm}

\includegraphics[width=0.75\linewidth]{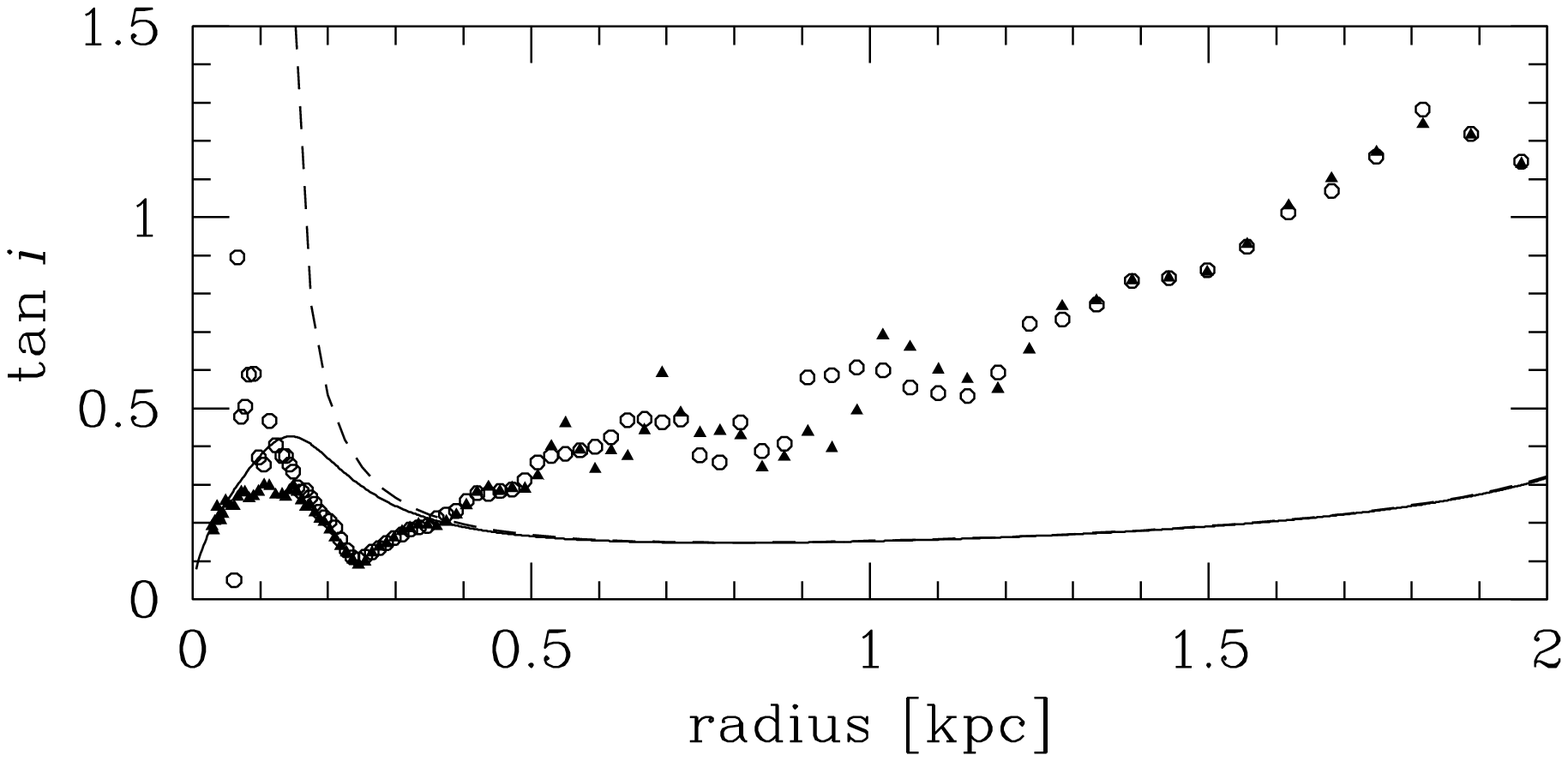}

\vspace{-33mm}

\caption{{\bf Top:} Snapshot of gas density (greyscale), and of 
$div^2 {\bf v}$ (for $div {\bf v} < 0$, contours, shock indicator) in model 
0S20r, at the time of 130 Myr, when the spiral shock reaches the inner grid 
boundary. The dashed circles mark the iILR at 0.13 kpc and the oILR at 2.3 
kpc. Overplotted are dotted circles of radii 40 pc, 100 pc, 250 pc, 500 pc, 
1 kpc and 2 kpc, in order to help in relating the amount of inflow in Fig.9 
to the observed morphology. Units on axes are in kpc.
{\bf Middle:} Radial density profile (dotted line), $div^2 {\bf v}$ for 
$div {\bf v} < 0$ (short-dashed line), and azimuthal velocity (solid line) 
with the rotation curve (long-dashed line) as a reference are plotted for 
the snapshot from the top panel as a function of radius along the line 
connecting the centre with the bottom-right corner of that panel. The 
velocity units are in \kms, the density and $div^2 {\bf v}$ units are 
arbitrary, but the same as in Fig.8.
{\bf Bottom:}  Tangens of the pitch angle $i$ of the shock, as indicated by
maxima of $div^2 {\bf v}$ in models 0S20r (open circles) and 8S20r (filled
triangles) plotted for the same time as the snapshot from the top panel. The
lines mark the linear prediction for an $m=2$ spiral in the potential of model 
8S20r (solid), and 0S20r (dashed).}
\label{f9}
\end{figure}

\subsection{Properties of the nuclear spiral}
I analyze properties of nuclear spirals in the central regions of a strongly 
barred galaxy using 2 hot-gas models: 0S20r and 8S20r. The early 
evolution of both models shows that as the time
passes, the nuclear spiral starts at the inner ends of the principal shocks 
and propagates inwards (Fig.7, top panel). However, the density enhancement 
related to it is very small (below 40\%), and the spiral is only seen as large 
$div^2 {\bf v}$, for negative $div {\bf v}$, which indicates the shock. 
Thus on the early stages 
of evolution, the principal shock in the bar gets extended inwards as a 
nuclear spiral shock. It has a number of properties that make it different 
from the density wave predicted by the linear theory:
\begin{itemize}
\item the strength of the shock does not drop significantly 
when its shape converts from straight to spiral; to the contrary, as can be
seen in the middle panel of Fig.7, the strength of the spiral shock (measured
by $div^2 {\bf v}$) at the radius of 400 pc, where 
it winds by $5\pi/4$ angle, is larger than that of the principal straight 
shock (at 1.5 kpc at this position angle);
\item the nuclear spiral in model 0S20r, having the iILR, does not stop at 
this resonance, but crosses it, and is propagating inwards, while in the linear
theory the wave does not extend beyond the resonance;
\item throughout the extent of the spiral shock, its pitch angle differs 
significantly from the linear prediction for both models 0S20r and 8S20r
(bottom panel of Fig.7), although model 0S05r indicates that it still 
increases with the sound speed in gas, as in the linear theory.
\end{itemize}

In both models 0S20r and 8S20r, at the simulation time about 130 Myr, the 
spiral shock reaches the inner boundary of the polar grid located at the 
radius of 20 pc. All plots in Fig.7 show characteristics of the models at
this moment. Due to the imposed reflective inner boundary condition, the
wave making the nuclear spiral reflects at this boundary and interferes with 
the incoming spiral wave. However the reflected wave geometrically diverges, 
and it perturbs the incoming wave, which converges on the centre, only at the 
innermost radii (see also Section 3.3). Note that the wave reflecting at the 
inner boundary may initially be weak, and not a shock, since $div^2 {\bf v}$ 
along the spiral decreases in the innermost parts of the galaxy at these 
early stages of evolution (Fig.7, top panel). 
This is consistent with the nuclear spiral in model 8S20r following the 
linear prediction for the pitch angle at the innermost radii below $\sim 50$
pc (Fig.7, bottom panel).

\begin{figure}
\centering

\vspace{-27mm}

\includegraphics[width=0.95\linewidth]{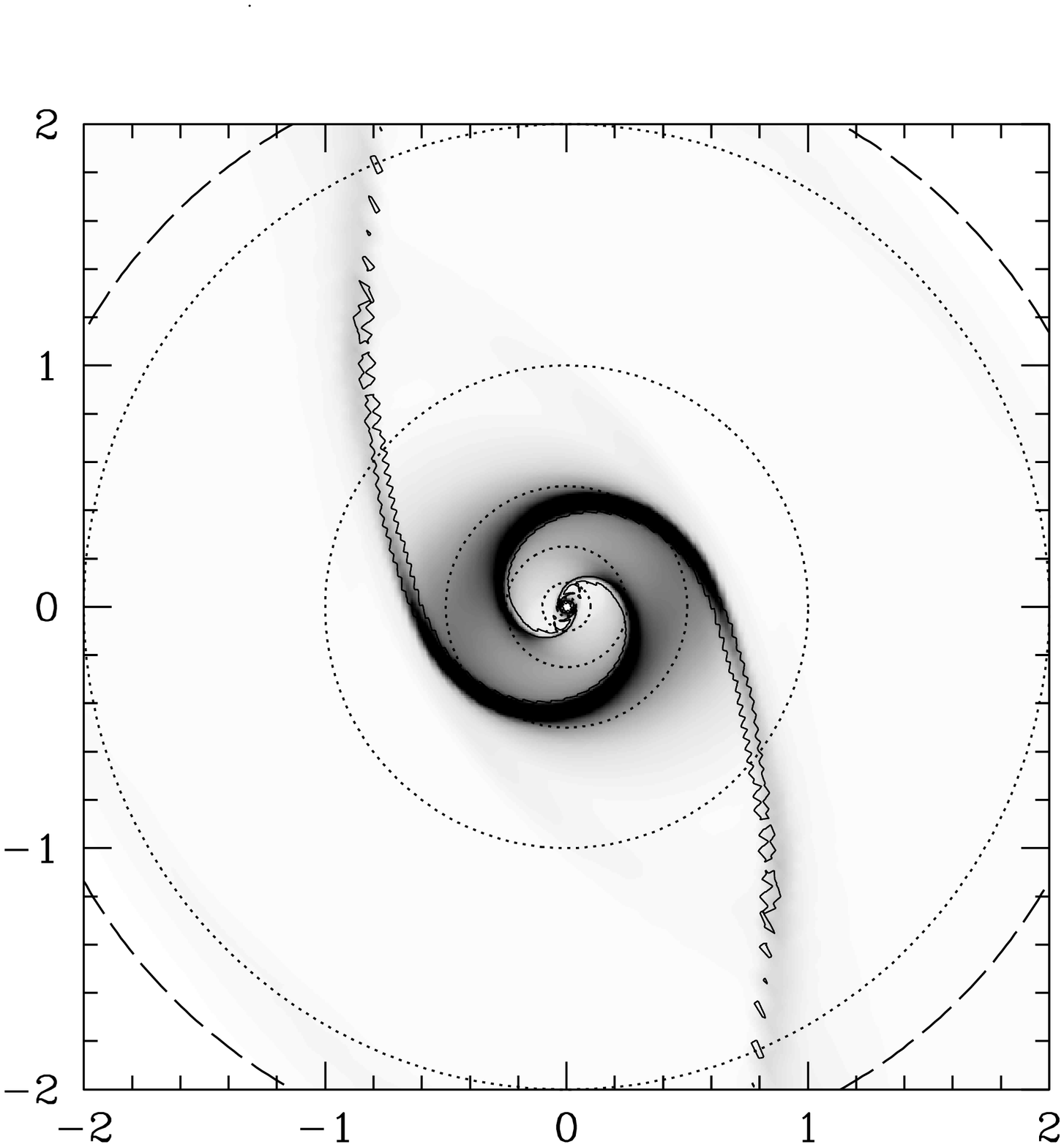}

\vspace{-7mm}

\includegraphics[width=0.75\linewidth]{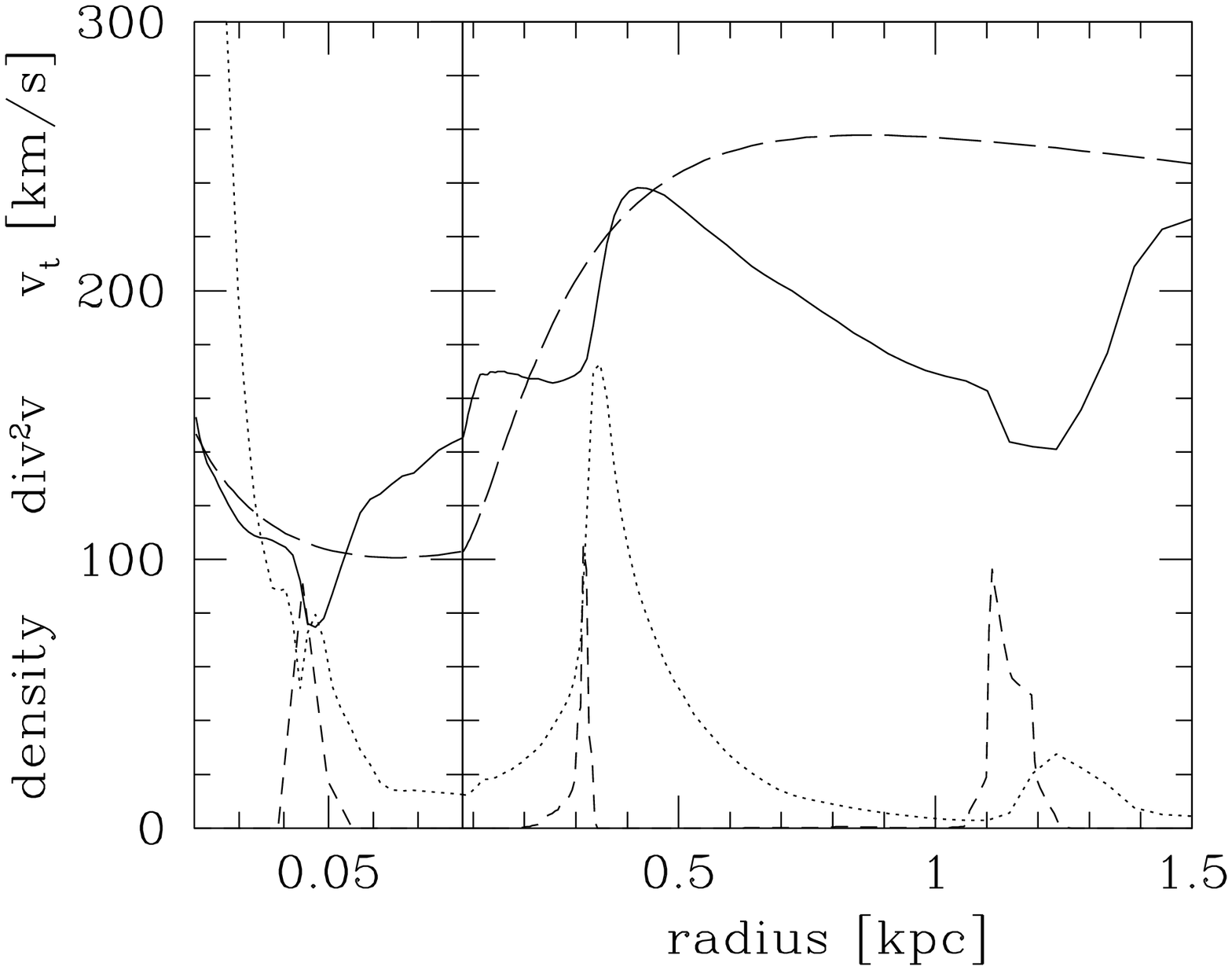}

\vspace{-14mm}

\includegraphics[width=0.75\linewidth]{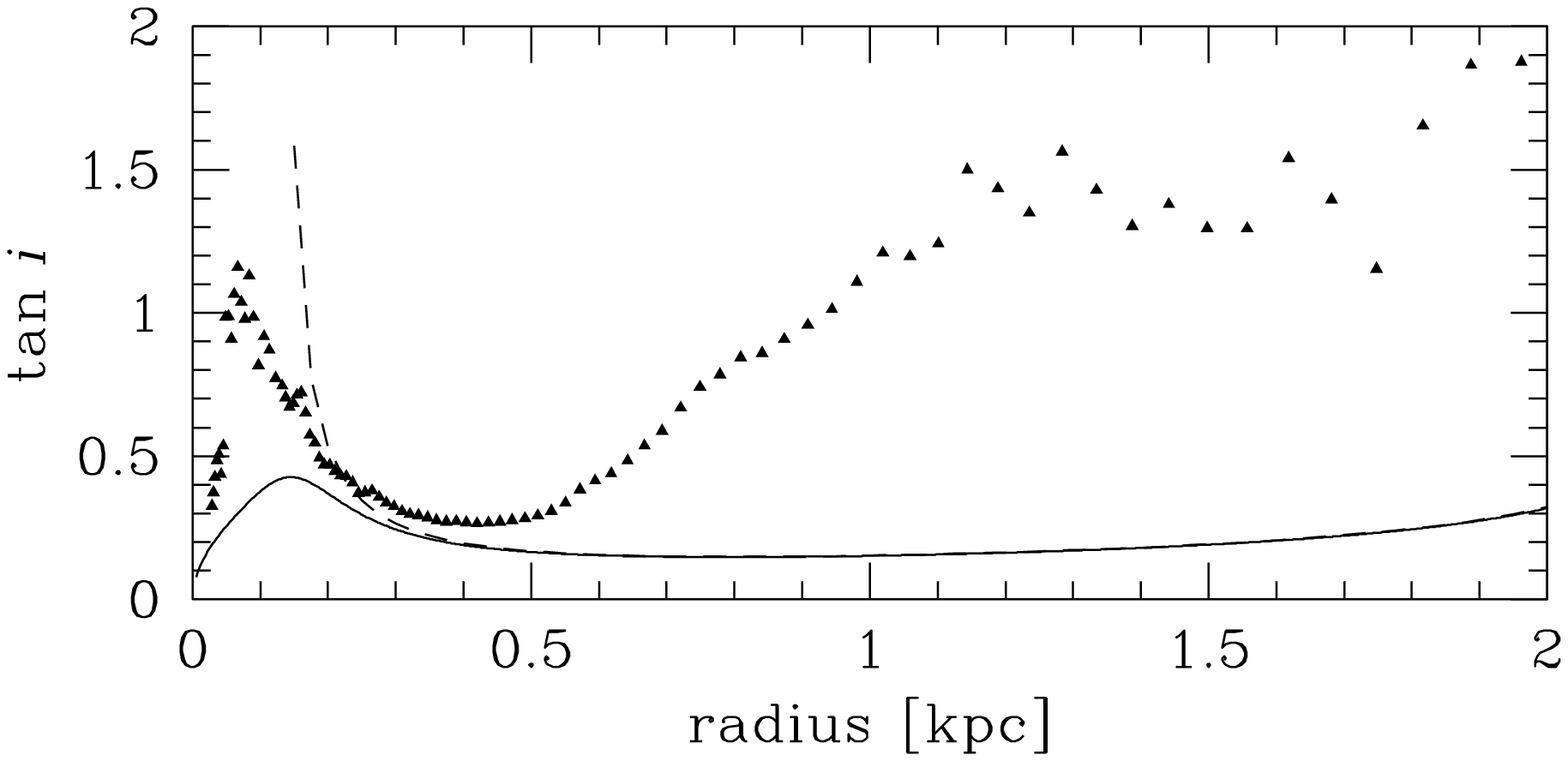}

\vspace{-33mm}

\caption{{\bf Top:} Same as in Fig.7, but for model 8S20r, at the time of 
0.5 Gyr, after the morphology of the flow has stabilized. The meaning of 
the circles, and the units are the same as in Fig.7.
{\bf Middle:} Same as in Fig.7, but for the snapshot from this top panel. To
achieve better resolution for the innermost features, the plot has two adjacent
parts drawn in different radial scales: the left one covers radii from 20 pc
to 80 pc, and the right one -- from 80 pc to 1.5 kpc. Velocity, density and
$div^2 {\bf v}$ units are the same as in Fig.7.
{\bf Bottom:}  Same as in Fig.7, but for the snapshot from this top panel.
Filled triangles mark the values measured in the hydrodynamical model, while
the solid line is the linear prediction. Note how the hydrodynamical shock
tends to follow the linear solution at radii 0.2 - 0.5 kpc, but still clearly
differs from it.}
\label{f10}
\end{figure}

After the reflection of the spiral wave from the inner boundary, its 
morphology quickly reaches a steady state, and it remains unchanged till
the end of the runs at 0.5 Gyr. In model 0S20r, the spiral shock recedes
from the centre after the reflection, and in the steady state it is confined
to the outside of the iILR. Characteristics of the flow in model 8S20r at 
the time when its appearance has stabilized are presented in Fig.8. Several 
interesting features of the flow can be observed:
\begin{itemize}
\item as can be seen in the middle panel of Fig.8, the strength of the shock 
is roughly the
same at 1.1 kpc, 0.3 kpc, and 0.05 kpc --- the first location is at the
principal straight shock, while the two last locations are at the nuclear
spiral shock; comparing middle panels of Fig.8 and Fig.7 one can see that the 
strength of the spiral shock at $0.3 - 0.4$ kpc has not changed throughout 
the run (models 0S20r and 8S20r do not differ much at radii that large);
\item variations in the profile of the tangential velocity (Fig.8, middle 
panel), which are much larger than in the model 8W20r with a weak oval 
(Fig.1, middle panel), also indicate that the departures from the circular 
rotation are nonlinear here, and indicative of a shock;
\item the structure of the shock is best resolved in the cut through the spiral
at 0.3 kpc (Fig.8, middle panel): regions of enhanced density (dotted line) 
occur
directly outside of the regions of large velocity convergence, which indicates
the shock (dashed line), with the contact between the two zones at 0.33 kpc;
for the trailing spiral it means that the density enhancement occurs
downstream from the shock;
\item contrary to the early stages of evolution (Fig.7, top panel), when
largest density concentration occurs around the principal shock, at later 
stages (Fig.8, top panel) it is located in the nuclear spiral; middle 
panels of Figs. 7 and 8 (drawn to scale) show that the peak
density increased between the early and late 
stage by a factor of about 2 in the principal shock at $1.1 - 1.9$ kpc, but 
in the nuclear spiral at $0.3 - 0.4$ kpc the rise is by a factor of more
than 20;
\item the pitch angle of the nuclear spiral (Fig.8, bottom panel) still 
differs from the linear prediction (it is persistently larger), although it 
shows similar trends: the linear wave theory proposed by Englmaier \& Shlosman
(2000) to explain the nuclear spirals in bars points out these trends, but the 
flow is nonlinear and literal application of the linear theory is not 
adequate here.
\end{itemize}

\subsection{Inflow in the spiral shock}
The inflow in the spiral shock has been determined analogously to that in
the weak nuclear spiral presented in section 3.4 (Fig.4). The evolution 
with time of mass accumulated within a number of radii for models 0S20r and 
8S20r is shown in the top panel of Fig.9. The difference between Figures 4 
and 9 is clear: there is strong inflow at virtually all radii in models with a 
bar, and especially in model 8S20r which includes a $10^8$\solm\ MBH in the
centre.

The circle of radius 4 kpc occurs slightly outside the 4:1 resonance in the
assumed potential, which is the outer limit of the straight principal shock
(see MTSS02), therefore inflow through this circle should be small. In fact,
the mass accumulated within this radius oscillates in both models 0S20r
and 8S20r between $7.2 \times 10^8$\solm\ and $8.9 \times 10^8$\solm\ 
throughout the run. 

The principal shock in the bar cuts through the circles of radii 2 kpc and 
1 kpc (see Fig.8, top panel), and largest inflow is expected here. 
In fact, in the period between the stabilizing of the
the morphology of the large-scale flow at 200 Myr, and the end of the run 
at 500 Myr, about $2.2 \times 10^8$\solm\ of gas crosses inwards through
each circle, which corresponds to the average inflow of 0.7 \solmyr (see
also bottom panels of Fig.9). Thus by the end of the simulation, the mass 
included in each of these circles increases several times compared to its 
initial value.

\begin{figure}
\centering
\includegraphics[width=0.95\linewidth]{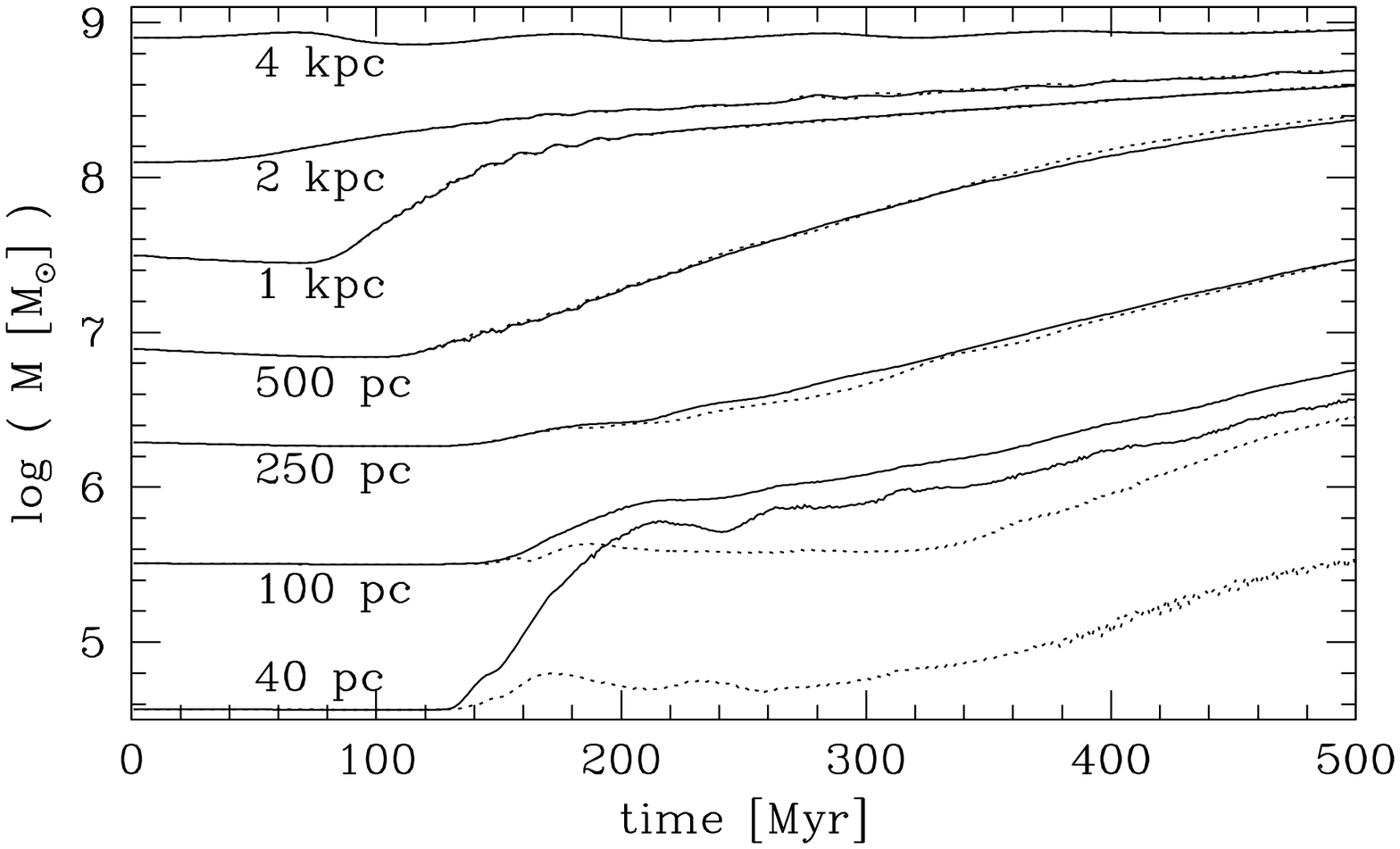}

\vspace{-3cm}

\includegraphics[width=0.75\linewidth]{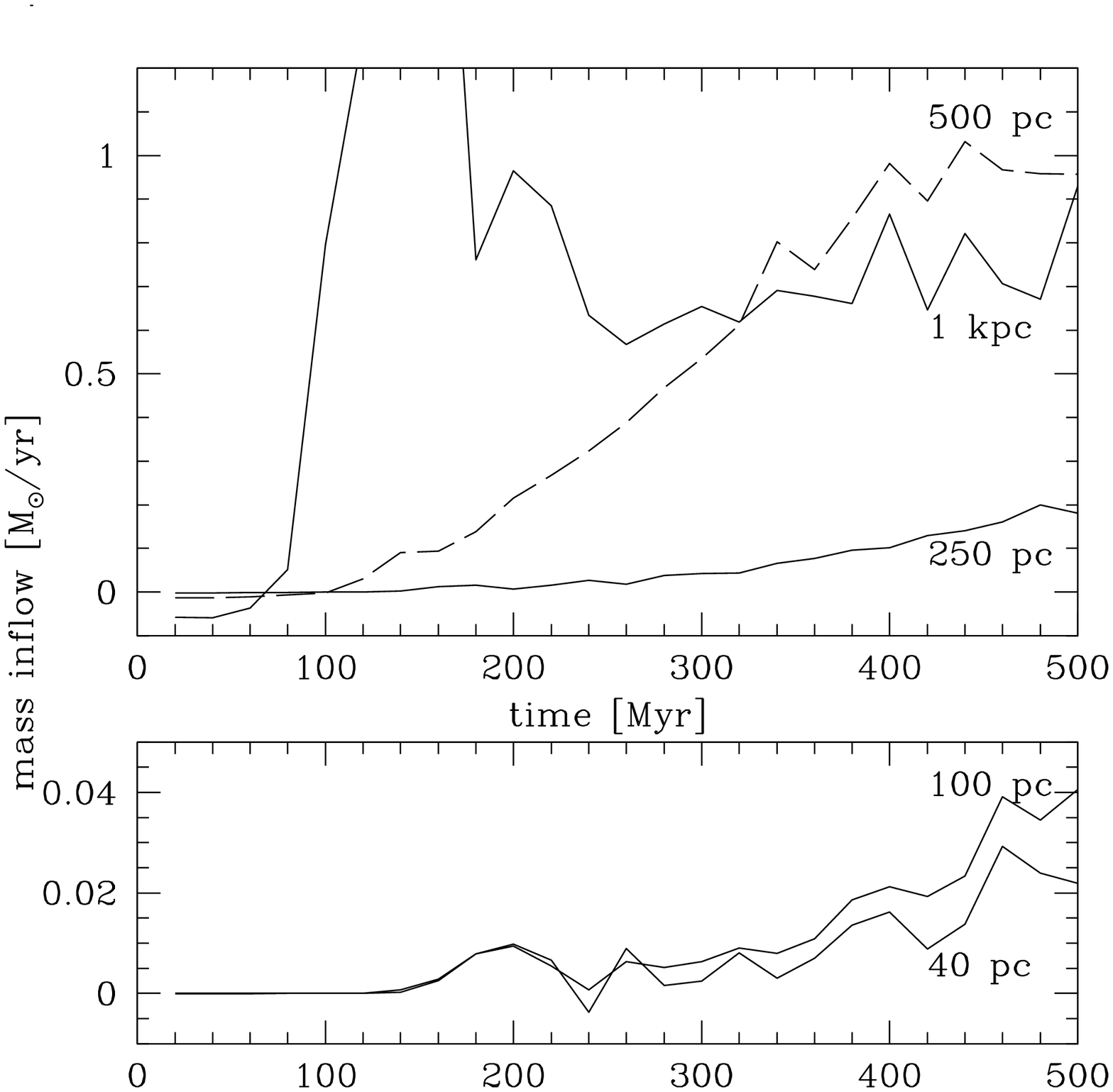}
\caption{{\bf Top:} Mass accumulated within various radii (indicated in the 
plot) as a function of time for model 8S20r (solid line) and 0S20r (dotted 
line).
{\bf Bottom:} Mass inflow averaged over 20-Myr intervals as a function of 
time in model 8S20r. The inflow is followed through circles of various 
radii indicated in the plot. Note the small, but not negligible inflow
triggered at the innermost radii in model 8S20r after the arrival of the 
spiral shock there (at about 130 Myr).}
\label{f11}
\end{figure}

The circles of radius 500 pc and smaller are located in the region of the
nuclear spiral shock. The mass accumulated within radii of 500 and 250 pc 
evolves 
identically in models 0S20r and 8S20r, therefore inflow at such radii most
likely is not influenced by the presence of a MBH in the galaxy centre.
Nevertheless, this inflow is considerable. It is so because the nuclear
spiral hosts a shock, whose nature is dissipative. This spiral shock,
like the principal shock in the bar, takes away angular momentum from
gas. However, the velocity jump in the
spiral shock is much smaller than that in the principal shock, and the
extraction of angular momentum is less efficient. The timescale of inflow
in the nuclear spiral is much longer than in the principal shock, and gas 
accumulates in the spiral, as is seen when comparing top panels of Figs. 7 
and 8. This results in extreme over-densities in the nuclear spiral (Fig.8,
middle panel): over hundred times larger than the initial density. 

Note that the
mechanism extracting angular momentum from gas in the the nuclear spiral
continues to work at the same efficiency per density unit. The increasing
density means increasing inflow. Thus gas from the inflow in the principal
shock gets collected on the nuclear spiral, and when its highest condensation 
moves inwards along this spiral towards a certain radius, the inflow at 
this radius increases. In the models
it can be seen on the example of the 500 and 250 pc radii. Top panel of Fig.8 
indicates that the maximum density in the spiral has already passed the 
radius of 500 pc, and keeps propagating inwards along the spiral. This is 
consistent with the inflow through this radius plotted in the bottom panels 
of Fig.9: it was constantly
increasing between 100 and 400 Myr, when the density peak in the spiral
was propagating inwards to reach the radius of 500 pc. When it reached
this radius, the amplitude of inflow stabilized. However, the inflow does 
not decay after that time, because the principal shock keeps the gas
supply open. Note that at the end of the run, the inflow in the nuclear
spiral at 500 pc is the same as that in the principal shock at 1 kpc,
which indicates that some kind of equilibrium has been established.

On the other hand, the density peak in the spiral does not reach the
radius of 250 pc within the simulation time (Fig.8, top panel), and inflow 
through this circle keeps increasing (Fig.9, middle panel). At the end of 
the run it reaches the value of 0.2\solmyr, 3.5 times smaller than that 
in the principal shock, and in the nuclear spiral at 500 pc.
However, the evolution of the models suggests that when the density 
peak reaches also this radius, the inflow will stabilize at the value
equal to that at the larger radii, and it will be the same for each smaller
radius, so that eventually a steady-state develops throughout the spiral,
with the inflow in the spiral equal to that in the principal shock.
However, it takes some 0.5 Gyr to establish such inflow at 500 pc, and likely
over 2 Gyr to establish it at 250 pc. Thus although in principle nuclear
spirals can cause strong inflow of about 1\solmyr at arbitrarily small radii,
it is uncertain whether such spirals exist for periods long enough, so that
the inflow has time to reach these small radii.

However, another mechanism of inflow in nuclear spirals generated by a
strong bar takes place at the innermost radii. In Section 3.4, I noticed
a period of inflow in the weak spiral related to the change in its innermost
morphology. It is best seen in Fig.4 as a change of mass accumulated within 
the radius of 40 pc. Similar increase within this radius is seen in model
8S20r after the spiral shock reaches the inner grid boundary at about 130 Myr
(Fig.9, top panel). Between that time and 200 Myr, the mass within the radius 
of 40 pc increases from $0.37 \times 10^5$ \solm\ to about $6 \times 10^5$. 
However, contrary to the weak spiral in model 8W20r, inflow in the spiral 
shock never ceases (Fig.9, bottom panel), but it rather increases with time, 
reaching the value of about 0.03 \solm/yr at the end of the run. Similar rate 
of inflow is present at the radius of 100 pc, and both rates similarly evolve 
in time. 

The inflow of $\sim 0.03$ \solmyr takes place only in the model with the MBH 
in the centre (8S20r). The top panel of Fig.9 indicates that the evolution of 
mass accumulated in the inner 40 and 100 pc is significantly different in model
0S20r without the MBH. In this model, after the spiral shock reaches the inner 
grid boundary, the mass enclosed within 40 pc increases initially by some 70\%,
but later it decreases, and oscillates around lower values. A quasi-monotonical
mass increase occurs after 300 Myr, but it is likely related to the first 
mechanism of inflow described above, which propagates inwards from larger 
radii. In any case, the mass enclosed within radius of 40 pc at the end of the
run in model 0S20r is $3.2 \times 10^5$ \solm, which corresponds to the average
inflow of 0.0015 \solmyr in the period between 300 and 500 Myr. This inflow 
is 20 times smaller than in model 8S20r with a MBH. 

\section{Nuclear spirals in double bars}
Hydrodynamical models of gas flow in dynamically possible double bars (each
bar supported by orbits calculated in this potential) were built by MTSS02. 
Already the orbital analysis (Maciejewski \& 
Sparke 2000) indicated that straight principal shocks cannot form in the 
inner bar in such systems, but gas should rather settle in rings elongated 
with the inner bar. Hydrodynamical models of MTSS02
confirmed these predictions, and evolutionary stars+gas models of Rautiainen 
et al. (2002) showed that gas settles on orbits calculated by Maciejewski
\& Sparke (2000). However, these models have been constructed for cold gas
only. 

When the outer bar is identical to that in the models analyzed in the previous 
section, it should by itself generate a nuclear spiral in the hot gas (see
model 8S20r, Fig.8, top panel). On the other hand, the orbital structure of 
the inner bar supports formation of gaseous rings, like the ring in model 
D05 in MTSS02, which is elongated with that bar. Again, the 
ring and the spiral should occur at the same location since the inner bar 
is confined within the ILR of the outer one. Thus also in the case of double 
bar, there is a competition between the orbital structure and the propagating 
wave.

In order to see what comes out of this competition, I built a model
of gas flow in the potential of a doubly barred galaxy identical to that
in the models of MTSS02, but this time for the hot gas. In this model, 
labeled 0D20o, the sound speed in gas is 20 \kms, and both bars
are being introduced simultaneously in the first 100 Myr of the run.

In the linear approximation outlined in Paper I, the solution is additive, and 
each independently rotating perturbation in the potential generates its own
spiral mode in gas, which propagates with a specific dispersion relation.
Note however that in the model considered here, the inner bar rotates with 
pattern speed 110 \kmskpc, therefore it has no ILR (see the top-left panel
of Fig.2 in Paper I). 
Thus this inner bar does not generate a nuclear spiral on its own. Only the 
outer bar, which has a wide ILR, generates a nuclear spiral in the inner 
kiloparsec, as model 0S20r indicates. 

\begin{figure}
\centering

\vspace{-12mm}

\includegraphics[width=0.48\linewidth]{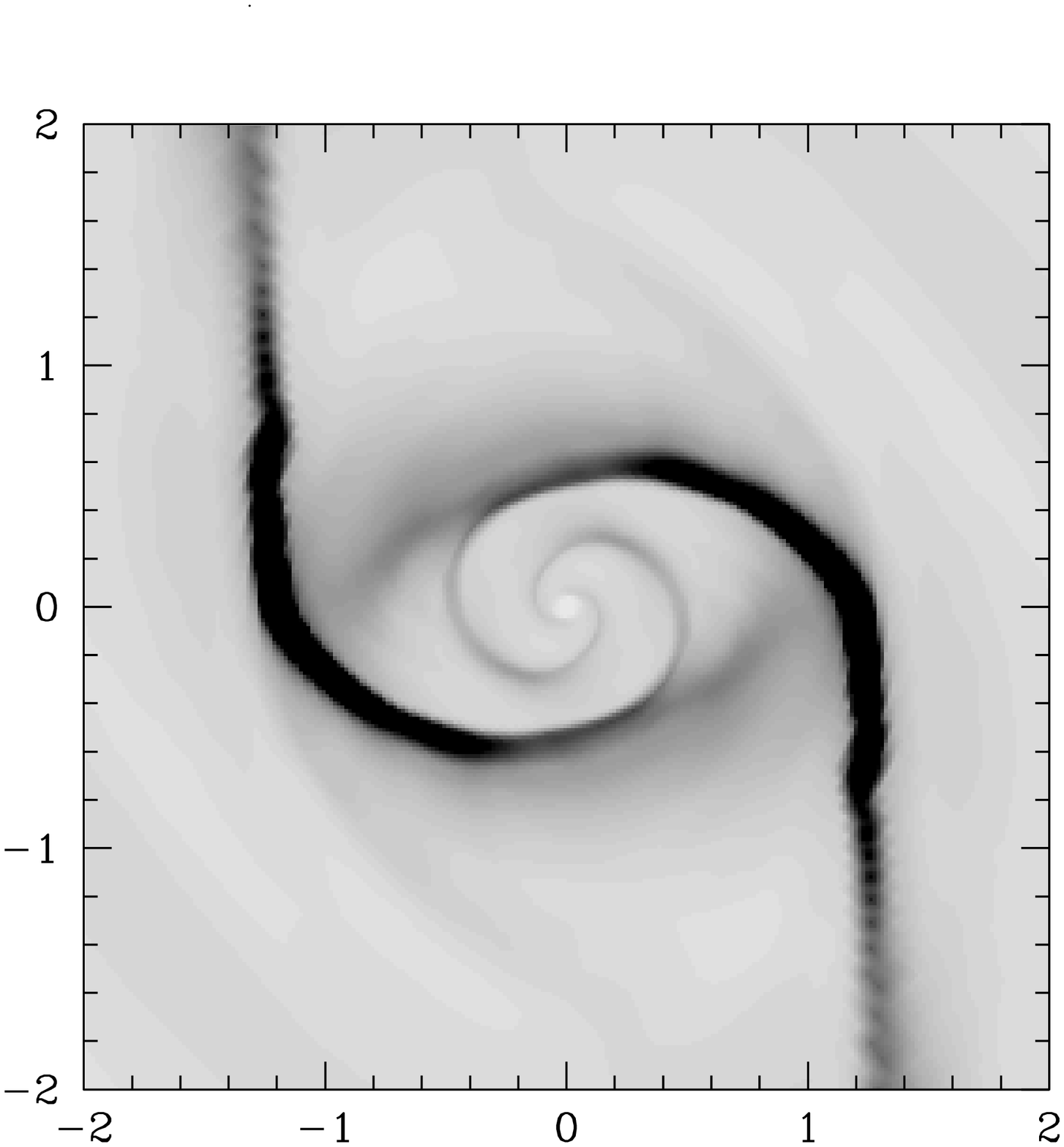}
\includegraphics[width=0.48\linewidth]{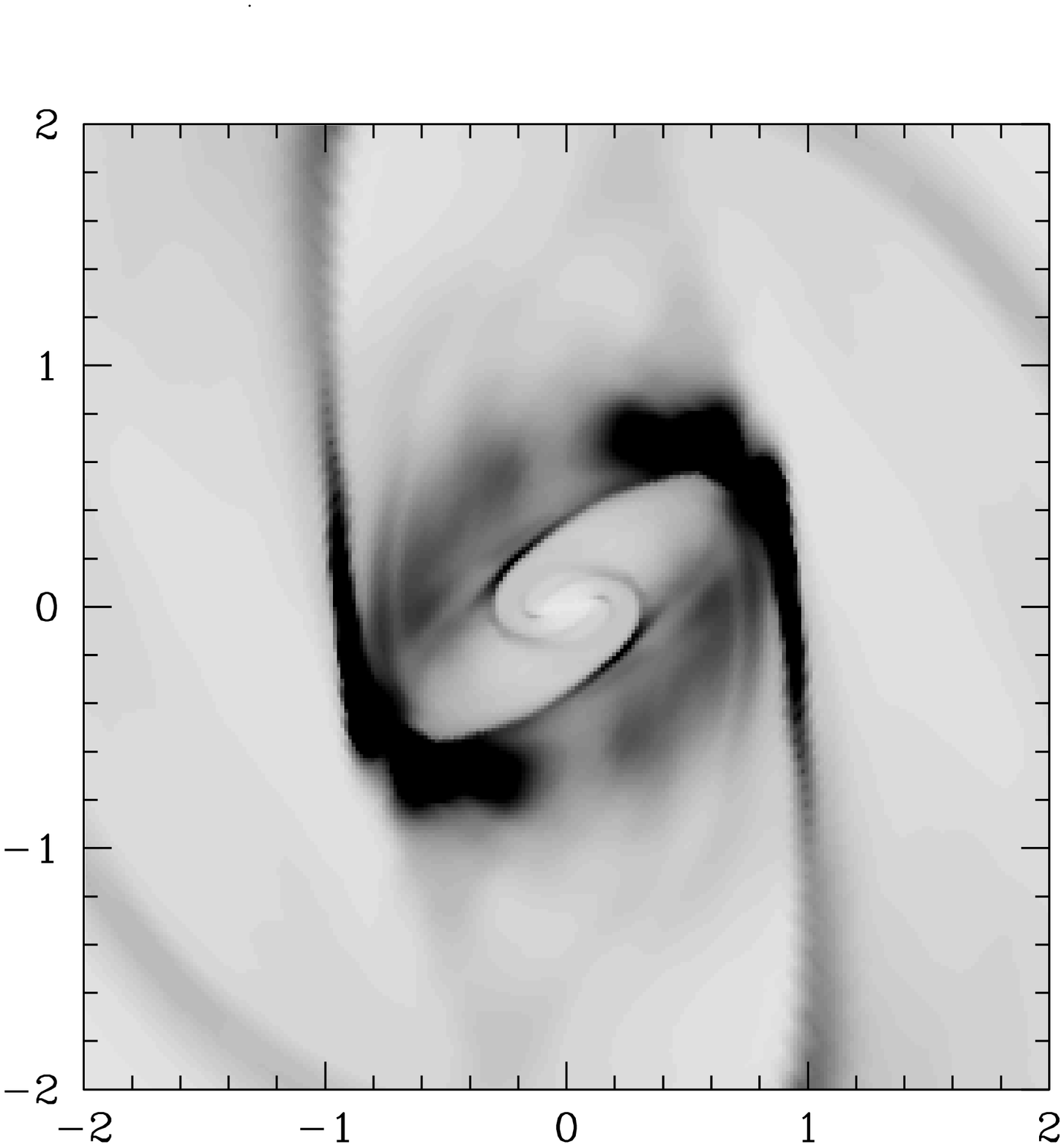}
\vspace{-5mm}
\caption{Snapshots of gas density in model 0D20o of hot gas in the 
gravitational potential of a doubly barred galaxy identical to that used in 
MTSS02, taken at 145 Myr ({\it left}), and 200 Myr ({\it right}). The outer 
bar is vertical, position of the inner bar can be deduced from oval density 
enhancements. Units on axes are in kpc.}  
\label{f12}
\end{figure}

Fig.10 shows two snapshots of gas density in model 0D20o for hot gas in a
doubly barred galaxy. Since the pitch angle of the spiral shock in such gas
is high, the spiral shock usually propagates out of the density enhancement
emerging from the straight principal shock. Therefore the nuclear spiral
propagates inwards in this doubly barred galaxy despite the action of 
the inner bar. It reaches the inner grid boundary at about 145 Myr (Fig.10, 
left panel). Although according to the linear theory
the curvature of the spiral does not change with time at any point 
in the frame rotating with the outer bar, the curvature of the trajectories
on which gas parcels move changes at a given point in this frame
with the rotation of the {\it inner bar}. The pitch angle of these trajectories
may become larger than the pitch angle of the spiral, causing discontinuities 
in the spiral shock propagating inwards. On later evolutionary stages, when the
spiral shock traps considerable amounts of gas around itself, its shape is
more influenced by the motion of the inner bar. At times, it may 
resemble straight principal shocks in the inner bar (Fig.10, right panel)
although such shapes are transient. Note that this structure is still caused 
by the outer bar, even if it resembles the principal shock in the inner bar, 
to which it may be wrongly ascribed. Also with time a broad ring forms 
around the inner bar, but with moderate over-density.

Because of the complexity of this  problem, detailed investigation of gas 
dynamics in nested bars requires further work, but this brief analysis
already returned some important information: nuclear spirals generated by
the outer bar in doubly barred systems can propagate inside the inner
bar. Thus the presence of nuclear spirals in galaxies does not exclude a 
cospatial coexistence of inner bars in the same galaxies. Moreover, nuclear
spirals can hide the presence of inner bars, as there is not much difference
in the gas kinematics between the systems displayed in the left panel of 
Fig.10 and in the top panel of Fig.7.

\section{Discussion}
\subsection{Morphology of nuclear spirals}
The number of galaxies with color- or structure maps of their central regions
has recently become large enough, so that first attempts of morphological 
classification have been made (Malkan et al. 1998, Martini et al. 2003a,b).
This second attempt seems to better reflect the characteristic structures
observed in galactic nuclei. There nuclear spirals are segregated into one 
of four classes: grand design, tightly wound, loosely wound and chaotic.
Here I attempt to link this classification to the morphology observed in the
models built in this paper.

Hydrodynamical models of gas flow in rotating potentials presented here
show that nuclear spirals are triggered even by small
asymmetries in the potential. There they propagate as weak density waves,
and they are not bracketed by straight principal shocks, as it is the
case in strong asymmetries induced by galactic bars. This should 
be expected, because the $x_1$ orbits supporting a weak oval asymmetry 
are round, with no cusps, and thus do not induce shocks in gas. 
Since there are no straight
shocks to join, the spirals can continue winding around the centre, 
closely following predictions of the linear theory. Thus tightly-wound 
nuclear spirals, which can propagate freely in weak asymmetries of the 
potentials, may be observationally associated with galaxies where the bar is
too weak to be detected. However, if the potential or conditions in gas
imply large pitch angle of the spiral wave in a given galaxy, then a loosely 
wound spiral will appear in a galaxy classified as unbarred.

On the other hand, in strong bars nuclear 
spirals rapidly unwind outwards to match the shape of the straight principal 
shock in the bar. Thus regardless of the underlying potential or velocity
dispersion in gas, nuclear spirals in strong bars are not likely to appear
observationally as tightly-wound spirals. This is consistent with the 
statistics of Martini et al. (2003 a,b): tightly wound spirals avoid 
barred galaxies.

Grand-design nuclear spirals require a strong driver which acts continuously 
over long time periods. In the statistics of Martini et al., they appear only 
in galaxies classified as barred. This implies that the galactic bar can serve 
as such a driver, and that there may be no other driver that fulfills the
criterion above. Tightly or loosely wound spirals in the classification of 
Martini et al. can be generated by a weak oval, or when they show clear 
discontinuities, by a passing perturbation in the potential (globular cluster 
or giant molecular cloud). With growing
discontinuities, one moves to the class of chaotic spirals, whose generation
mechanism is likely different (acoustic noise, see Elmegreen et al. 1998).

\subsection{Gas kinematics. Feeding of the AGN.}
Recent observational statistics of the central morphology in a sample
of active galaxies accompanied by a control sample (Martini et al. 2003 a,b)
indicates that nuclear spirals occur with comparable frequency in active
and non-active galaxies. On the other hand, models presented in this paper 
show that nuclear spirals generated by a strong bar take the form of 
shocks in gas and trigger moderate gas inflow onto the central MBH, while
nuclear spirals generated by a weak oval do not cause the inflow. I propose
that what determines the inflow is not the driver, but the nature
of the spiral. If it is a shock, then it is likely to trigger inflow. If
it is not, inflow will not occur. Note that the morphology of the spiral
shock departs from the linear prediction (Fig.8, bottom panel) in the sense 
that the pitch angle of the spiral is larger than in the linear theory. Thus
spiral shocks that do not appear as grand-design spirals, are likely to 
be observationally classified as {\it loosely wound} spirals. 

In the full sample of Martini et al. (2003a), grand-design and loosely wound
spirals occur in 60\% of active galaxies, and only in 23\% of inactive
ones. This difference is statistically significant, and it may indicate
that although not all nuclear spirals are fueling the AGN, some spirals
most likely do it. However, morphological considerations are not sufficient 
to verify this hypothesis, and observations of gas kinematics in the spirals
is needed. Clear departures from circular motion are expected in spiral
shocks (Fig.8), but not in weak density waves that do not trigger inflow
(Fig.1).

It became recently generally accepted that all galaxies may host a MBH at their
centres (see e.g. Kormendy \& Gebhardt 2001 for a recent review), and there are
attempts to measure the mass of this MBH from the gas kinematics around it
(e.g. Macchetto et al. 1997, Bower et al. 1998, Maciejewski \& Binney 2001).
This method can be derailed by non-circular gas motions in the nuclear discs,
especially when they exhibit spiral structure. Spiral shocks can strongly 
perturb the velocity field (Fig.8). However, 
if tightly wound nuclear spirals in fact correspond to models where the spiral 
is a weak density wave, then gas flow in such a spiral is almost 
circular (Fig.1), and methods based on gas kinematics should yield a reliable 
MBH mass here.

\subsection{Leading spirals, nuclear rings}
According to the linear theory outlined in Paper I, nuclear spirals generated
at the OLR and at the oILR are trailing, and hydrodynamical models built in 
this paper well reproduce the trailing spirals related to the oILR. These 
spirals 
propagate inwards, as expected. However, the linear theory also predicts 
formation of a {\it leading} spiral at the iILR that propagates outwards.
Such combinations of a leading spiral inside a trailing one are very unusual
in galaxies, with the only familiar example being NGC 6902 (Grosb{\o}l 2003). 
In the models
presented here leading spirals do not form, even when the iILR is present.
One reason for it may be the extent of the trailing spiral generated at the
oILR, which propagates inwards to the vicinity of the iILR in model 0W20, 
and even past it in model 0S20 of a spiral shock. This may suppress formation
of the leading spiral. However, in model 0W05, where the trailing spiral gets
damped not far inwards from the oILR, leading spiral does not form either.
Some explanation here may come from the applications of the non-linear theory
of density waves developed by Yuan \& Kuo (1997). It allows to investigate
the effect of viscosity on the nuclear spirals. In the results presented by
Yuan, Lin \& Chen (2003) it can be seen that the leading spiral forms only 
for high viscosity. Viscosity in the hydrodynamical code used to build models 
in this paper is very low, and this may be the reason for the absence of the 
leading spiral. On the other hand, its absence in the observed nuclei of 
galaxies may indicate low effective viscosity in the ISM, much lower than 
what the quality of the code often imposes on the available hydrodynamical 
models.

Another feature promoted by numerical models and definitely under-abundant
in the observed galaxies are nuclear rings. Only 2 nuclear rings are found
in the sample of 43 Seyfert galaxies observed by Pogge \& Martini (2002),
and an eye-examination of the larger sample of 123 galaxies (Martini 2003a)
picks up a dozen of nuclear rings. In Section 4.1 I showed that nuclear rings 
form as an effect of interaction between the wave-nature of the principal
shock in bar, and the orbital structure there. This interaction creates 
conditions favourable to the formation of nuclear rings when the velocity 
dispersion in gas is low. In fact, low velocity dispersion in gas is assumed 
in models that form nuclear rings (e.g. Piner et al. 1995, Regan \& Teuben 
2003). However, velocity dispersion in the inner discs of spiral galaxies is 
likely to be higher (Englmaier \& Gerhard 1997, Elmegreen et al. 1998), which
is consistent with the observed frequency of nuclear spirals that is higher 
than that of nuclear rings. Thus studies of nuclear rings may only partially
reflect gas dynamics in centres of galaxies. In particular, stagnation of gas 
inflow in the bar caused by these rings is not a general evidence against the 
possibility that bar-related inflow can occur at radii typical for rings.
Nuclear rings make only one type of nuclear gas flow, and another type, 
nuclear spirals, may extend bar-related inflow to the innermost regions of 
galaxies.

\section{Conclusions}
In this paper I analyzed high-resolution hydrodynamical models of gas flow
in nuclear spirals generated in the gaseous disc by a rotating potential.
Nuclear spirals form naturally even if the asymmetry in the potential is very
small, with the maximal ratio of radial to tangential force (the $Q_T$ 
parameter, Combes \& Sanders 1981) about 0.01. Thus asymmetries in galaxies 
often to weak to be detected observationally, like a weak triaxiality of the 
bulge, may be sufficient to generate nuclear spirals. Models with weak 
asymmetry in the potential well conform to the linear prediction, while the 
nature of nuclear spirals in strong bars considerably differs from what 
is predicted by the linear density-wave theory. 

Models of galaxies with weak ovals indicate that nuclear spirals form 
even if the asymmetry in the potential
is too weak to generate straight shocks along its major axis. In such
potentials, nuclear spirals are not forced to unwind rapidly outwards to
match the straight shocks: they can follow predictions of the linear
density-wave theory for longer, and wind more tightly than in the presence 
of a strong bar. This is consistent with the recent statistical analysis
by Martini et al. (2003a,b) which finds that tightly wound nuclear spirals
rather avoid galaxies that are barred. The smooth continuation of the nuclear 
spiral into a four-arm spiral seen in the models with weak ovals indicates 
that the extent of such tightly-wound spirals may not be a good indicator
of the location of the ILR in galaxies without a clear bar. Nuclear spirals
in weak ovals are not efficient in transporting gas from kiloparsec- to
parsec- scale, but some inflow in the innermost parsecs of the galaxy occurs 
during formation of such a spiral in models without an iILR (with a
central MBH). Since these nuclear spirals can naturally re-appear as a
response to the driver, gas dumped each time onto the MBH can maintain a
weak nuclear activity.

In strong bar the nuclear spiral has the nature of a shock in gas. The spiral
shock is less tightly wound than what the linear theory predicts (compare Fig.1
to Fig.8). This may 
suggest correspondence between spiral shocks and loosely wound spirals in the
classification of Martini et al. (2003a). Hydrodynamical models built in this
paper show that such spiral shocks trigger gas inflow, although of different 
nature than that in the straight principal shock in the bar. In the outer
regions of the nuclear spiral, the inflow timescale is longer than in the
principal shock bracketing it from outside. Therefore gas initially
accumulates there, but with time it is transported inwards along the spiral.
The inflow rate at density peaks along the spiral equals that in the
principal shock. Another inflow mechanism is present in the innermost tens
of parsecs of the nuclear spiral in the presence of a central MBH: after 
the initial dumping of matter onto the centre, common with the models for
weak ovals, the mass inflow does not stop, but continues at a steady rate of
up to 0.03 \solmyr. Local Seyfert galaxies require mass accretion rates of 
$\sim 0.01$ \solmyr (e.g. Peterson 1997), therefore the inflow rate in
the models presented here is sufficient to feed luminous local Active
Galactic Nuclei, and the feeding can continue over long timescales.
An observational support for this mechanism comes from the fact that
when one groups together grand-design nuclear spirals (explicitly linked 
to the bars) and loosely wound spirals, they appear considerably more often in
active than in non-active galaxies (Martini et al. 2003a,b).

Nuclear spirals are more common in galaxies than nuclear rings. Which of 
these two will be triggered by a barred potential depends on the interplay 
between the post-shock gas condensations, which 
tend to follow the lowest-energy orbits, and the shock, whose inner shape
adheres to the rules for wave-propagation. In the ISM with low velocity
dispersion, the shock is damped in gas condensation, and a nuclear ring
forms. When velocity dispersion in the ISM is high, the shock propagates 
away from gas condensation, gets strengthened, and continues inwards as a 
nuclear spiral. Higher frequency of nuclear spirals than of nuclear rings
in galaxies favours ISM with high velocity dispersion in centres of disc
galaxies.

Secondary inner bars in barred galaxies do not halt the propagation of 
nuclear spirals inwards. Thus nuclear spirals can co-exist with inner
bars in galaxies. Moreover, they can mask the presence of inner bars in
galactic nuclei.

\section*{Acknowledgments}
The CMHOG hydrodynamical code was written by James Stone, and I am grateful
to him for the right to use it. I would like to thank Lia Athanassoula, Peter
Erwin, and Paul Martini for useful discussions and for comments on this
paper. Helpful comments of the anonymous referee provided a good guidance in
improving the clarity of this paper. 
I acknowledge the post-doctoral fellowship from Osservatorio Astrofisico 
di Arcetri, where most of this research has been done. I am grateful to
Instytut Astronomiczny Uniwersytetu Wroc{\l}awskiego for its hospitality 
during writing this paper.

\end{document}